\documentclass[a4paper,11pt]{article}
\usepackage{jheppub}
\usepackage[T1]{fontenc}
\usepackage{amsmath,amssymb,amsfonts,graphicx,bm,bbm}
\usepackage{array}
\usepackage{float}
\usepackage{multirow,multicol}
\usepackage{subfigure}
\usepackage{titlesec}

\providecommand{\tabularnewline}{\\}

\newcolumntype{x}[1]{
{\centering\hspace{0pt}}p{#1}}

\newcommand{\GeV}{{\rm ~GeV}}
\newcommand{\TeV}{{\rm ~TeV}}
\newcommand{\pb}{{\rm ~pb}}
\newcommand{\fb}{{\rm ~fb}}
\newcommand{\ab}{{\rm ~ab}}
\newcommand{\invfb}{{\rm ~fb^{-1}}}
\newcommand{\invab}{{\rm ~ab^{-1}}}

\newcommand{\BR}[1]{{{\rm BR}\left(#1\right)}}

\newcommand{\vareps}{\varepsilon}

\newcommand{\I}{\rm 1\kern-.24em l} 
\newcommand{\Tr}{\mathop{\rm Tr}}

\newcommand{\Gam}[1]{\Gamma\left(#1\right)}
\newcommand{\mStar}{M_{V^*}}
\newcommand{\confirm}{\textcolor{black}}
\title{\boldmath QCD Corrections to Pair Production of Type III Seesaw Leptons at Hadron Colliders}

\author{Richard Ruiz}

\affiliation{Pittsburgh Particle physics, Astrophysics, and Cosmology Center {\rm(Pitt-PACC)},\\
Department of Physics $\&$ Astronomy, University of Pittsburgh, Pittsburgh, PA 15260, USA}
\affiliation{Institute for Particle Physics Phenomenology {\rm(IPPP)},\\ 
Department of Physics, Durham University, Durham, DH1 3LE, UK\footnote{Address as of October 2015.}}

\emailAdd{richard.ruiz@durham.ac.uk}

\abstract{If kinematically accessible, hadron collider experiments provide an ideal laboratory for the direct production of heavy lepton partners in Seesaw models.
In the context of the Type III Seesaw Mechanism, the $\mathcal{O}(\alpha_s)$ rate and shape corrections are presented
for the pair production of hypothetical, heavy $SU(2)_L$ triplet leptons in $pp$ collisions at $\sqrt{s}=$ 13, 14, and 100 TeV.
The next-to-leading order (NLO) $K$-factors span, approximately, $K^{NLO}=1.1 - 1.4$ for both charged current and neutral current processes 
over a triplet mass range $m_T = 100~\text{GeV}-2~\text{TeV}$. 
Total production cross sections exhibit a $^{+5\%}_{-6\%}$ scale dependence at 14 TeV and $\pm1\%$ at 100 TeV.
The NLO differential $K$-factors for heavy lepton kinematics are largely flat, suggesting that na\"ive scaling by the total $K^{NLO}$ is reasonably justified.
The resummed transverse momentum distribution of the dilepton system is presented at leading logarithmic (LL) accuracy.
The effects of resummation are large in TeV-scale dilepton systems.
Discovery potential to heavy lepton pairs at 14 and 100 TeV is briefly explored:
At the High-Luminosity LHC, we estimate a $4.8-6.3\sigma$ discovery potential maximally for $m_T = 1.5-1.6~\text{TeV}$ after 3000 fb$^{-1}$.
With 300 (3000) fb$^{-1}$, there is $2\sigma$ sensitivity up to $m_T = 1.3-1.4~\text{TeV}~(1.7-1.8~\text{TeV})$ in the individual channels.
At 100 TeV and with 10 fb$^{-1}$, a $5\sigma$ discovery can be achieved for $m_T=1.4-1.6~\text{TeV}$.
Due to the factorization properties of Drell-Yan-type systems, the fixed order and resummed calculations reduce to convolutions over tree-level quantities.}

\arxivnumber{1509.05416}

\begin{document}
\begin{flushright}
PITT-PACC-1513
\end{flushright}
\hfill
\maketitle
\flushbottom


\section{Introduction}
The origin of sub-eV neutrino masses is a central issue in particle physics.
As right-handed neutrinos do not exist in the Standard Model (SM), which thus predicts massless neutrinos, 
new particles are necessary to explain neutrino masses~\cite{Ma:1998dn},
e.g., 
gauge singlet fermions in the Type I~\cite{Minkowski:1977sc,Mohapatra:1979ia,Yanagida:1979as,GellMann:1980vs,Schechter:1980gr,Shrock:1980ct} Seesaw Mechanism, 
or scalar and fermionic SU$(2)_L$ triplets in the Types II~\cite{Magg:1980ut,Cheng:1980qt,Lazarides:1980nt,Mohapatra:1980yp} 
and III~\cite{Foot:1988aq} scenarios.
Searches for these degrees of freedom constitute an important component of hadron collider programs; 
see Refs.~\cite{Barger:2003qi,Atre:2009rg,Chen:2011de,Deppisch:2015qwa} and references therein.
Furthermore, the maturity of the formalism underlying QCD corrections in hadron collisions, 
which are required for predicting accurate production rates and distribution shapes, 
readily permit their application to beyond the SM (BSM) processes.

\begin{figure}[!t]
\includegraphics[width=0.96\textwidth]{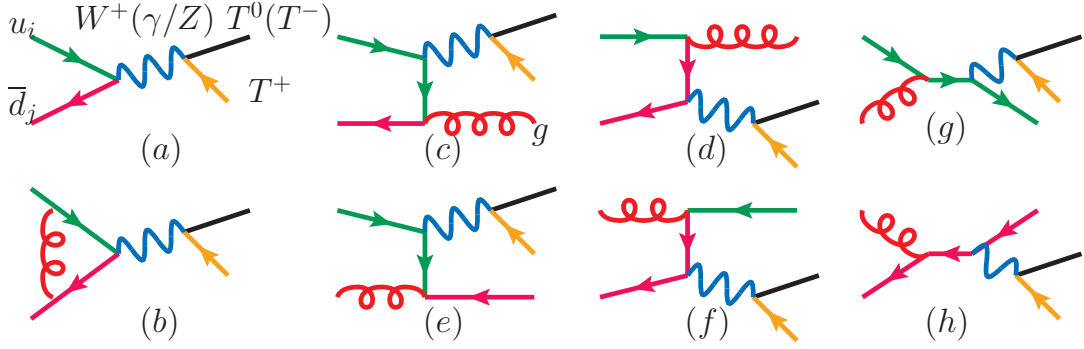}
\caption{
(a) Leading order Feynman diagram for $pp\rightarrow T^0T^\pm$ and $T^+T^-$ production.
(b-h) $\mathcal{O}(\alpha_s)$ corrections.
Drawn with JaxoDraw~\cite{Binosi:2003yf}.}
\label{fig:feynman}
\end{figure}

For heavy Seesaw partners with sub-TeV masses, the dominant hadron collider production mode is through the Drell-Yan (DY) 
charged current (CC) and neutral current (NC) processes~\cite{Keung:1983uu,Pilaftsis:1991ug,Datta:1993nm,Han:2006ip,Rizzo:1981dm,Rizzo:1983zz,Gunion:1996pq}, 
shown in figure~\ref{fig:feynman}(a).
For TeV-scale systems and above, the $W\gamma$ fusion channel becomes dominant~\cite{Datta:1993nm,Dev:2013wba,Alva:2014gxa}.
A catalog of resonant Seesaw partner production modes in hadron collisions is given in Ref.~\cite{Datta:1993nm}.

In the Type I Seesaw, production cross section is known at next-to-leading order (NLO) in QCD~\cite{Ruiz:2015gsa}
and estimated at next-to-next-to-leading order (NNLO) 
via a $K$-factor\footnote{The $N^mLO$ $K$-factor is defined as $K = \sigma^{N^mLO} / \sigma^{LO}$, 
where $\sigma^{LO}$ is the lowest order $(m=0)$, or  Born, cross section and $\sigma^{N^mLO}$ is the N$^m$LO-corrected cross section.}~\cite{Alva:2014gxa}.
For the Type II case, rates are known at NLO~\cite{Muhlleitner:2003me,Sullivan:2002jt},
NLO with next-to-leading logarithm (NLL) recoil and threshold resummations~\cite{Jezo:2014wra}, and automated at NNLO~\cite{Gavin:2010az,Gavin:2012sy}.

Pair production of heavy Type III Seesaw leptons has, until now, been evaluated only to leading order (LO) accuracy.
For $m_T = 100\GeV-2\TeV$, we report the NLO $K$-factors:
\begin{eqnarray}
\confirm{ 1.17-1.37} &\quad\text{at}\quad& \sqrt{s}=13\TeV,\\
\confirm{ 1.17-1.36} &\quad\text{at}\quad& \sqrt{s}=14\TeV,\\
\confirm{ 1.14-1.29} &\quad\text{at}\quad& \sqrt{s}=100\TeV,
\end{eqnarray}
with scale uncertainty of $\confirm{^{+5\%}_{-6\%}}$ at 14 TeV and $\confirm{\pm1\%}$ at 100 TeV,
and are comparable to other DY-type processes in Seesaw models.
The NLO differential $K$-factors\footnote{The differential $N^mLO$ $K$-factor with respect to observable $\hat{\mathcal{O}}$ is 
defined as $K_{\hat{\mathcal{O}}} = \left(d\sigma^{N^mLO}/d\hat{\mathcal{O}}\right)  / \left(d\sigma^{LO}/d\hat{\mathcal{O}}\right)$.}
for heavy lepton kinematics are largely flat for TeV-scale $m_T$, suggesting that na\"ive scaling by the total $K^{NLO}$ is reasonably justified.

In this study, production rates of TeV-scale Type III Seesaw lepton pairs
at $\mathcal{O}(\alpha_s)$ accuracy are presented for $pp$ collisions at $\sqrt{s} = 13,~14,$ and $100\TeV$. 
Differential distributions at NLO and NLO with leading logarithm (LL) resummation of TeV-scale lepton kinematics are presented for the first time at 14 TeV.
The fixed order (FO) calculation is carried out via phase space slicing (PSS)~\cite{Fabricius:1981sx,Kramer:1986mc,Baer:1989jg,Harris:2001sx}.
The calculation of the dilepton system's transverse momentum, $q_T$, 
follows the Collins-Soper-Sterman (CSS) formalism~\cite{Collins:1981uk,Collins:1981va,Collins:1984kg}.
This text continues in the following order:
In section~\ref{sec:theory}, we summarize the Type III Seesaw model and comment on experimental constraints.
The PSS and CSS formalisms are briefly introduced in section~\ref{sec:methodology}.
Due to the factorization properties of DY-type systems, the fixed order and resummed results reduce to convolutions over tree-level quantities;
technical details are relegated to appendices \ref{app:NLO} and \ref{app:LL}.
Results are reported in section~\ref{sec:results}. We summarize and conclude in section~\ref{sec:summary}.

\section{Type III Seesaw Mechanism}~\label{sec:theory}
\subsection{Model Lagrangian}
The Type III Seesaw~\cite{Foot:1988aq} generates tree-level neutrino masses via couplings to SU$(2)_L$ triplet leptons with zero hypercharge.
In terms of Pauli matrices $\sigma^a$, the left-handed (LH) fields are denoted by
\begin{equation}
 \Sigma_L = \Sigma^a_L\sigma^a = 
\begin{pmatrix} \Sigma^3_L & \sqrt{2} \Sigma^+_L \\ \sqrt{2}\Sigma^-_L & -\Sigma^3_L \end{pmatrix},
\quad \Sigma^\pm_L \equiv \frac{\Sigma^1_L \mp i\Sigma^2_L}{\sqrt{2}},
\end{equation}
where $\Sigma^\pm_L$ have U$(1)_{\rm EM}$ charges $Q=\pm1$,
and the right-handed (RH) conjugate fields are
\begin{equation}
 \Sigma_R^c = \begin{pmatrix} \Sigma^{3 c}_R & \sqrt{2} \Sigma^{-c}_R \\ \sqrt{2}\Sigma^{+c} & -\Sigma^{3 c}_R \end{pmatrix}.
\end{equation}
Chiral conjugates are related by $\psi_R^c \equiv (\psi^c)_R = (\psi_L)^c$, where $\psi_{L/R} \equiv P_{L/R}\psi = \frac{1}{2}(1\mp\gamma^5)\psi$.

For a single generation (but generalizable to more), the model's Lagrangian is
\begin{eqnarray}
 \mathcal{L}_{\rm Type~III} &=&  \mathcal{L}_{\rm SM}  + \mathcal{L}_{T} + \mathcal{L}_Y,
\end{eqnarray}
where $\mathcal{L}_{\rm SM}$ is the SM Lagrangian, the triplet's covariant derivative and mass are given by
\begin{equation}
 \mathcal{L}_{T} = 
 \frac{1}{2}\Tr\left[\overline{\Sigma_L}i\not\!\!D\Sigma_L\right] 
-\left(\frac{m_T}{2}\overline{\Sigma_L^3}\Sigma_R^{3c} + m_T\overline{\Sigma^-_L}\Sigma^{+c}_R + \text{H.c}\right), 
\end{equation}
and the SM LH lepton $(L)$ and Higgs $(\Phi)$ doublet fields couple to $\Sigma_L$ via the Yukawa coupling
\begin{equation}
 \mathcal{L}_Y = -y_T \overline{L} ~\Sigma_R^c ~i\sigma^2 \Phi^* + \text{H.c.}, \quad v = \sqrt{2}\langle\Phi\rangle \approx 246\GeV.
\end{equation} 
Dirac masses are then spontaneously generated after electroweak symmetry breaking (EWSB):
\begin{eqnarray}
 \mathcal{L}_Y \overset{\rm EWSB}{\longrightarrow}
 \mathcal{L}_Y &= &
 -\frac{y_T}{\sqrt{2}}(v+h)\overline{\nu_L}\Sigma^{3c}_R - y_T(v+h)\overline{e_L}\Sigma^{+c}_R + \text{H.c.},
\end{eqnarray}
leading to the neutral fermion mass  matrix
\begin{eqnarray}
\frac{y_T v}{\sqrt{2}}\overline{\nu_L}\Sigma^{3c}_R + \frac{m_T}{2}\overline{\Sigma_L^3}\Sigma_R^{3c}  + \text{H.c.}  = 
\frac{1}{2}
\begin{pmatrix} \overline{\nu_L} & \overline{\Sigma^3_L} \end{pmatrix}
\begin{pmatrix} 0 & y_T v/\sqrt{2} \\ y_T v/\sqrt{2} & m_T \end{pmatrix}
\begin{pmatrix} \nu_R^c \\ \Sigma^{3c}_R \end{pmatrix}  + \text{H.c.}
\label{eq:SeesawIIIMixing}
\end{eqnarray}
The Seesaw mechanism proceeds by supposing $m_T\gg y_T\langle\Phi\rangle$, leading to light/heavy mass eigenvalues
\begin{equation}
 m_{\rm light} \approx \frac{y_T^2 v^2}{2 m_T} \quad\text{and}\quad m_{\rm heavy} \approx m_T.
 \label{eq:SeesawIIIMasses}
\end{equation}
Thus, tiny neutrino masses follow from mixing with heavy states, 
whereby light (heavy) mass eigenstates align with the doublet (triplet) gauge states.
For $y_T$ comparable to the electron's SM Yukawa, sub-eV $m_{\rm light}$ 
can be explained by sub-TeV $m_T$, a scale within the LHC's kinematic reach.

We combine the $\Sigma$ fields and their conjugates into physical Dirac and Majorana fields:
\begin{eqnarray}
 \tilde{T}^- \equiv \Sigma^-_L + \Sigma^{+c}_R,\quad 
 \tilde{T}^+ \equiv \tilde{T}^{-c}, \quad 
 \tilde{T}^0 \equiv \Sigma^3_L + \Sigma^{3c}_R. 
\end{eqnarray}
In the gauge basis and in terms of $\tilde{T}$, the triplet interaction Lagrangian $\mathcal{L}_{T}$ is written as
\begin{eqnarray}
 \mathcal{L}_{T}^{\rm Gauge~Basis} &=& 
 \overline{\tilde{T}^-}\left(i\partial_\mu\gamma^\mu - m_T\right)\tilde{T}^- 
 + \frac{1}{2}\overline{\tilde{T}^0}\left(i\partial_\mu\gamma^\mu - m_T\right)\tilde{T}^0\nonumber\\
 &-& \overline{\tilde{T}^-}\left( e A_\mu\gamma^\mu  + g\cos\theta_W Z_\mu\gamma^\mu	 \right)\tilde{T}^- 
 -g \overline{\tilde{T}^-} W^-_\mu\gamma^\mu \tilde{T}^0 - g\overline{\tilde{T}^0}W^+_\mu\gamma^\mu \tilde{T}^-.
\end{eqnarray}
Our aim is to report the $\mathcal{O}(\alpha_s)$ corrections to heavy lepton pair production, 
which are independent of the mixing between the gauge states $\tilde{T}$ and mass states 
$T,~\ell~(\ell=e,\mu,\tau),$ and $\nu_m~(m=1,2,3)$.
For the remainder of the study, we generically denote the mixing as $Y$ and $\vareps$, and write 
\begin{equation}
 \tilde{T}^\pm = Y~T^\pm + \vareps~ \ell^\pm, \quad
 \tilde{T}^0   = Y~T^0   + \vareps~ \nu_m, \quad  \vert Y\vert\sim \mathcal{O}(1), \quad \vert\vareps\vert\ll1.
\end{equation}
The resulting interaction Lagrangian in the mass eigenbasis relevant to our study  is 
\begin{eqnarray}
 \mathcal{L}_{T}^{\rm Mass~Basis} &\ni& 
 \overline{T^-}\left(i\partial_\mu\gamma^\mu - m_T\right)T^- + 
 \frac{1}{2}\overline{T^0}\left(i\partial_\mu\gamma^\mu - m_T\right)T^0\nonumber\\
 &-& \overline{T^-}\left( e Y A_\mu\gamma^\mu  + g\cos\theta_W Y Z_\mu\gamma^\mu \right)T^- 
 -g Y \overline{T^-} W^-_\mu\gamma^\mu T^0 - g Y^* \overline{T^0}W^+_\mu\gamma^\mu T^-.
 \label{eq:SeesawLag}
\end{eqnarray}

\subsection{Constraints on Type III Seesaw Lepton Production}
For a review of constraints and phenomenology of the Type III Seesaw, 
see Refs.~\cite{Franceschini:2008pz,delAguila:2008cj,Arhrib:2009mz,Li:2009mw,AguilarSaavedra:2009ik,Bandyopadhyay:2010wp,Chen:2011de}.
\begin{itemize}
  \item \textbf{Collider Production and Decay:}  CMS experiment searches for $T^0T^\pm$ production and decay into
  $\ell^+\ell^-\ell^{'\pm}\not\!\!E_T$ have restricted the cross section and branching ratio to~\cite{CMS:2015mza}
  \begin{equation}
   \sigma(pp\rightarrow T^0T^\pm)\times {\rm BR}(T^0T^\pm\rightarrow \ell^+\ell^-\ell^{'\pm}\not\!\!E_T) < 12~\text{fb} ~\text{at the 95\% C.L.}
  \end{equation} 
  For equal doublet-triplet mixing among the SM leptons, this translates to the bound
  \begin{equation}
  m_T < 278~\GeV~\text{at the 95\% C.L.}
 \end{equation}
  With 20.3 fb$^{-1}$ of 8 TeV LHC data, searches carried out by the ATLAS experiment for 
  $T^0T^\pm\rightarrow W^\pm W^\pm \ell^\mp\not\!\!E_T \rightarrow 4j\ell^\mp\not\!\!E_T$ excludes~\cite{Aad:2015cxa}, depending on mixing parameters,
  \begin{equation}
   m_T < 325-540\GeV  ~\text{at the 95\% C.L.}
  \end{equation}
\end{itemize}

Throughout this study, we take $T^0$ and $T^\pm$ to be mass degenerate.
Electroweak (EW) corrections at one loop induce a mass splitting of $\Delta m_T \approx 160$ MeV 
for $m_T>100\GeV$~\cite{Pierce:1993gj,Ibe:2006de,Arhrib:2009mz}, and is thus negligible.
For differential distributions, we use representative mass
\begin{equation}
m_T = 500\GeV.
\end{equation}
As in the LO case, the total partonic and hadronic cross sections at NLO and NLO+LL factorize into a product of
the mixing parameter $\vert Y\vert$ and a mixing-independent ``bare'' cross section $\sigma_0$:
\begin{equation}
 \sigma^{NLO}(pp\rightarrow T\overline{T}) = \vert Y\vert^2 \times \sigma^{NLO}_0(pp\rightarrow T\overline{T}).
 \label{eq:BareDef}
\end{equation}
Therefore, we express our results in terms of $\sigma_0$ and do not choose any particular $\vert Y\vert$.
Furthermore, factorization implies that that total and differential NLO $K$-factors are independent of $\vert Y\vert$.

\section{Heavy Lepton Pair Production at $\mathcal{O}(\alpha_s)$ in Hadron Collisions}\label{sec:methodology}
Here we outline the PSS~\cite{Fabricius:1981sx,Kramer:1986mc,Baer:1989jg,Harris:2001sx}
and CSS~\cite{Collins:1981uk,Collins:1981va,Collins:1984kg} formalisms, which we use to calculate the processes
\begin{eqnarray}
 p ~p \rightarrow	 W^{\pm *} 		\rightarrow T^0 ~ T^\pm  \quad\text{and}\quad
 p ~p \rightarrow	 \gamma^*/Z^* 	\rightarrow T^+ ~ T^-,
 \label{eq:SeesawCCNC}
\end{eqnarray}
at NLO in QCD and the transverse momentum $q_T$  of the dilepton systems at LL.
With PSS and CSS, the inclusive NLO and NLO+LL results factorize and can be expressed in terms of tree-level, partonic cross sections.
Such technical details are given in appendices \ref{app:NLO} and \ref{app:LL}.
For simplicity, we generically denote processes in Eq.~(\ref{eq:SeesawCCNC}) and their radiative corrections  by 
\begin{eqnarray}
  p ~p \rightarrow  T ~\overline{T} \quad\text{and}\quad   p ~p \rightarrow  T ~\overline{T} ~j.
\end{eqnarray}
We note that these corrections are not unique but are well-known and general for the production of any $SU(2)_L$ triplet color-singlet, 
e.g.,~\cite{Altarelli:1979ub,Baer:1997nh}.
However, unlike previous studies, we investigate the $\mathcal{O}(\alpha_s)$ effects on the kinematic distributions of TeV-scale leptons.

\subsection{Phase Space Slicing}
To evaluate $T\overline{T}$ production at NLO, we follow the usual procedure: 
evaluate virtual and radiative corrections to the LO process in $d=4-2\vareps$ dimensions;
collect soft divergences, which cancel exactly; collect collinear divergences, which cancel partially;
and subtract residual collinear poles from parton distribution functions (PDFs).

For an $n$-body LO process, we divide, or \textit{slice}, the phase space of its $(n+1)$-body correction into soft and collinear kinematic regions.
For radiation energy $E_j$, partonic c.m.~energy $\sqrt{\hat{s}}$, and small dimensionless cutoff parameters $\delta_S,~\delta_C$,
a volume of the $(n+1)$-body phase space is \textit{soft} if
\begin{equation}
  E_j < \frac{\sqrt{\hat{s}}}{2}\delta_S.\label{eq:pssSoftCutDef}
\end{equation}
For partonic-level invariant masses and momentum transfers
\begin{equation}
\hat{s}_{ik} = (p_i + p_k)^2 \quad\text{and}\quad \hat{t}_{ik} = (p_i - p_k)^2,
\end{equation}
where indices $i,k$ run over initial- and final-state momenta, a region of phase space is \textit{collinear} if
\begin{equation}
 \hat{s}_{ik},~\vert \hat{t}_{ik}\vert < \hat{s} ~\times~ \delta_C.\label{eq:pssColCutDef}
\end{equation}
A volume is \textit{hard}~(\textit{non-collinear}) if not soft (collinear).
Exact choices of $\delta_S,~\delta_C$ do not matter: 
dependences on $\delta_S,~\delta_C$ cancel for sufficiently inclusive processes~\cite{Harris:2001sx}. 
However, so soft and collinear factorization remain justified, one needs
\begin{equation}
\delta_C\ll\delta_S\ll1.
\end{equation}
The hard-non-collinear $T\overline{T}j$ process is then finite everywhere and 
given by 
\begin{eqnarray}
  & & \sigma_{(3)}(pp\rightarrow T ~\overline{T} ~j ~X) =  \nonumber\\ & & 
\sum_{a,b=q,\overline{q}',g}
 \int_{\tau_0}^1 d\xi_1 \int^1_{\tau_0/\xi_1}d\xi_2 
 \left[f_{a/p}(\xi_1,\mu_f^2)f_{b/p}(\xi_2,\mu_f^2) + (1\leftrightarrow2)\right]\hat{\sigma}^{B}(ab\rightarrow T ~\overline{T} ~j),
 \label{eq:3Body}
\end{eqnarray}
where $\hat{\sigma}^{B}$ is the Born-level $\overline{T}T j$ partonic cross section.
For $a,b\in\{q,\overline{q}',g\}$ with $q\in\{u,d,c,s\}$,  
the PDF $f_{a/p}(\xi_i,\mu_f^2)$ is the likelihood of parton $a$ carrying away longitudinal momentum fraction $\xi_i$ from  proton $p$ 
evolved to a factorization scale $\mu_f$.
The c.m.~beam energy $\sqrt{s}$ and partonic c.m.~energy are related by $\hat{s} = \xi_1 \xi_2 s$, and 
we denote the threshold at which $T\overline{T}$ production occurs by $\tau_0$:
\begin{equation}
 \tau_0 = \min\frac{\hat{s}}{s} = \frac{(m_T + m_{\overline{T}})^2}{s}.
\end{equation}

In the soft/collinear limits, amplitudes for soft, soft-collinear, 
and hard-collinear radiation factorize into divergent expressions proportional to the (color-connected) Born amplitude.  
The poles are grouped with virtual corrections and the PDFs, resulting in a finite expression given by~\cite{Harris:2001sx} 
\begin{eqnarray}
& & \sigma_{(2)}(pp\rightarrow T ~\overline{T} ~X) =  \nonumber\\ & & 
\sigma^{HC} + 
 \sum_{a,b=q,\overline{q}}
 \int_{\tau_0}^1 d\xi_1 \int^1_{\tau_0/\xi_1}d\xi_2 
 \left[f_{a/p}(\xi_1,\mu^2_f)f_{b/p}(\xi_2,\mu^2_f) + (1\leftrightarrow2)\right]\hat{\sigma}_{(2)}(ab\rightarrow T ~\overline{T}),
 \label{eq:2Body}
\\
 & & \hat{\sigma}_{(2)} = \hat{\sigma}^{B} + \hat{\sigma}^{V} + \hat{\sigma}^{S}  + \hat{\sigma}^{SC},\label{eq:pss2BNonHC}
\end{eqnarray}
were $\hat{\sigma}^{B}$ is the Born-level $T\overline{T}$ partonic cross section, 
$\hat{\sigma}^{V}$ is its $\mathcal{O}(\alpha_s)$ virtual correction, 
$\hat{\sigma}^{S}$ and $\hat{\sigma}^{SC}$ are the soft and soft-collinear radiation terms, and 
$\sigma^{HC}$ is the hard-collinear radiation correction.
Inclusive triplet lepton production at NLO is now reduced to a sum of two- and three-body processes:
\begin{eqnarray}
 \sigma^{\rm NLO}(pp\rightarrow T ~\overline{T} ~X) &=& 
 \sigma_{(2)}(pp\rightarrow T ~\overline{T} ~X) + \sigma_{(3)}(pp\rightarrow T ~\overline{T} ~j ~X).
\label{eq:SigmaNLO}
 \end{eqnarray}
 
\begin{figure}[!t]
\subfigure[]{	\includegraphics[width=0.48\textwidth]{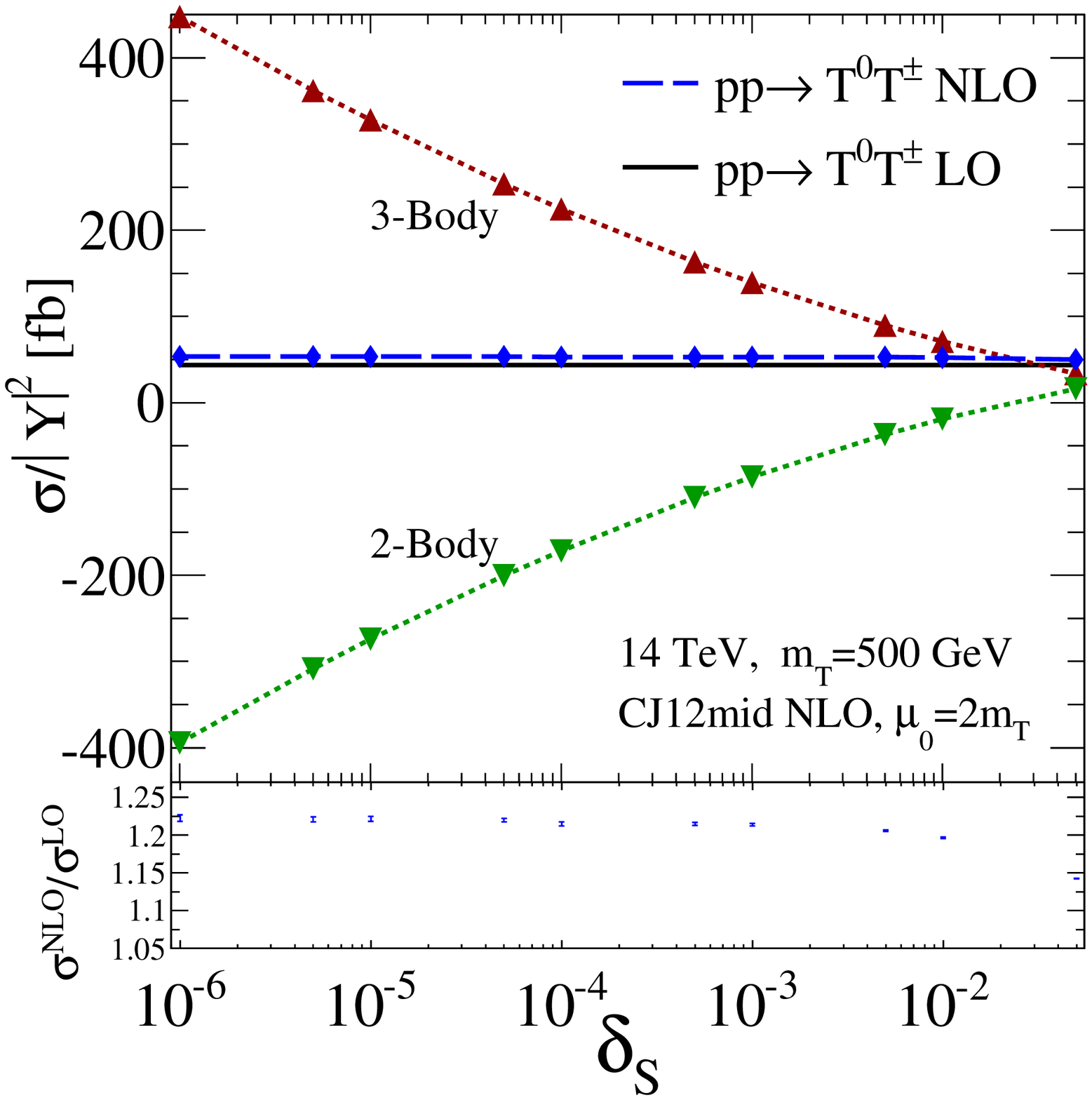}	\label{fig:cc_XSec_vs_Soft}	}
\subfigure[]{	\includegraphics[width=0.48\textwidth]{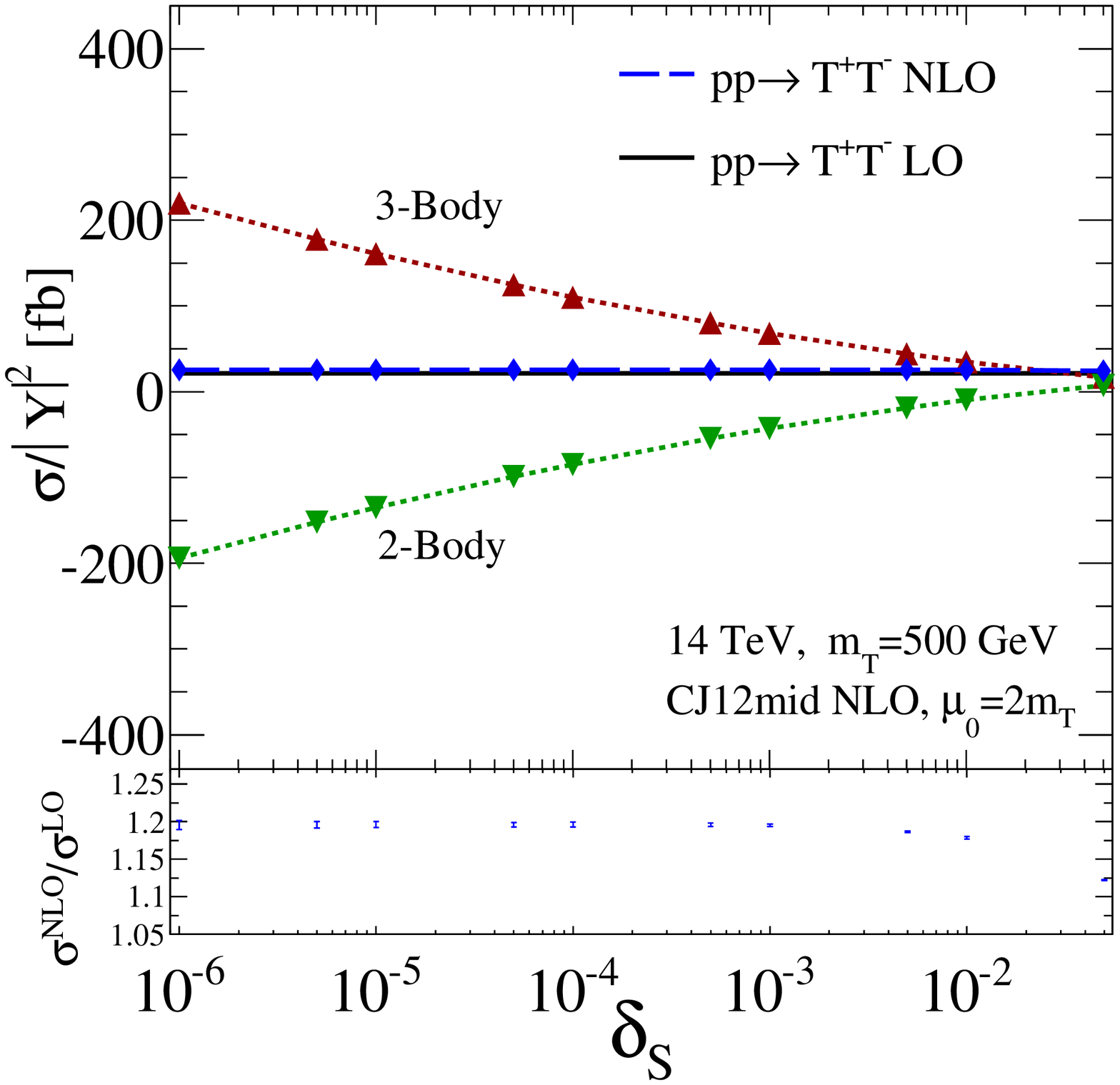}	\label{fig:nc_XSec_vs_Soft}	}
\caption{The 14 TeV LHC (a) $T^0T^\pm$ and (b) $T^+T^-$ production cross section, divided by $\vert Y\vert^2$, 
at LO (solid) and NLO (dash) as a function of $\delta_S$
with the two- (dot-upside down triangle) and three-body (dot-triangle) NLO contributions.
Panel: Ratio of NLO to LO with Monte Carlo uncertainty.}
\label{fig:softResults}
\end{figure}

Using the inputs from section~\ref{sec:results} and representative triplet mass $m_T = 500\GeV$,
in figures~\ref{fig:cc_XSec_vs_Soft} and \ref{fig:nc_XSec_vs_Soft} we show as a function of soft cutoff $\delta_S$
the 14 TeV $T^0T^\pm$ and $T^+T^-$ bare production cross sections at LO (solid) and NLO (dash) [Eqs.~(\ref{eq:SigmaNLO})]
with the two-body (dot-upside down triangle) [Eqs.~(\ref{eq:2Body})] and three-body (dot-triangle) [Eqs.~(\ref{eq:3Body})] NLO contributions.
In the panel, we show the NLO $K$-factor with Monte Carlo uncertainty. 
The negative value of $\sigma_{(2)}$ for $\delta_S \lesssim \mathcal{O}(0.5)$ is due to large hard-collinear PDF subtractions.
The NLO result is insensitive to $\delta_S$ for $\delta_S\lesssim1-3\times10^{-3}$, 
reflecting the large but fine cancellation of Eqs.~(\ref{eq:3Body}) and (\ref{eq:2Body}).
For example: for $\delta_S = 1\times10^{-5}$, the three- and two-body calculations are approximately 
$\confirm{+10.2}$ and $\confirm{-9.0}$ times the LO cross section.
For $\delta_S\gtrsim3\times10^{-3}$, the $K$-factor plummets, 
indicating the importance of terms proportional to powers of $\delta_S$ in the two-body expression, and hence a breakdown of soft/collinear factorization.

\subsection{Collins-Soper-Sterman Transverse Momentum Resummation}\label{sec:cssMethod}
As a color-singlet process, colored initial state radiation (ISR) is the dominant contribution to the $T\overline{T}$ system's $q_T$ spectrum.
For dilepton invariant mass $\mStar$, the distribution has the power series
\begin{equation}
\frac{d\sigma(pp\rightarrow T ~\overline{T} ~X)}{dq_T^2} = \sum_{k=1} ~ A_k ~ \alpha_s^k(\mStar^2) ~
 \log^{(2k-1)}\left(\frac{\mStar^2}{q_T^2}\right), ~ \quad A_k \sim \mathcal{O}(\alpha^2),\label{eq:resPowerSeries}
\end{equation}
and indicates a breakdown of the perturbative description in the $q_T^2/\mStar^2\rightarrow0$ limit.
The leading $\alpha_s \log(\mStar^2/q_T^2)$ logarithm in FO calculations is only reliable when~\cite{Collins:1984kg}
\begin{equation}
 \log\frac{\mStar}{\Lambda_{\rm QCD}} \gg 1 \quad\text{and}\quad \log^2\frac{\mStar}{q_T} \lesssim \log\frac{\mStar}{\Lambda_{\rm QCD}}, \quad 
 \Lambda_{\rm QCD}=0.2\GeV.
 \label{eq:ResEstimation}
\end{equation}
That is, $\alpha_s(\mStar^{2})$ must be perturbative and the scales associated with the process must be comparable.
As the Seesaw mass scale is pushed higher~\cite{Aad:2015cxa,CMS:2015mza}, 
so too does the scale at which artificially large logarithms appear.
For $m_T= 500\GeV~(1\TeV),$ FO predictions breakdown at $q_T \sim 55~(95)\GeV$,
and resummation of recoil logarithms become necessary to describe $q_T$ below this threshold.

Fortunately, as $q_T/\mStar\rightarrow0$, gluon radiation factorizes.
This permits one to reorganize, sum, and exponentiate large logarithms in Eq.~(\ref{eq:resPowerSeries}), 
resulting in an all-orders expression in terms of the Born process.
The resummed distribution with respect to $q_T^2, ~M^2_{V^*},$ and dilepton rapidity $y$,
is ~\cite{Collins:1984kg}
\begin{equation}
 \frac{d\sigma^{\rm Resum}(pp\rightarrow T ~\overline{T} ~X)}{{d\mStar^2 ~dy ~dq_T^2}} = 
 \frac{1}{4s}\int^\infty_0 db^2 ~J_0(b q_T) ~\times W ~\times ~d\hat{\sigma}^{B}(q\overline{q}\rightarrow T ~\overline{T}).
 \label{eq:ResumExpression}
\end{equation}
The integral is over the impact parameter $b$ and is the Fourier transform of $q_T$;
the zeroth order Bessel function $J_0$ emerges as a simplification.
$W$ expands to a perturbative and non-perturbative set of universal Sudakov form factors, and a process-dependent luminosity weight $\tilde{W}$:
\begin{equation}
 W = e^{-S_{NP}}~ e^{-S_P} ~  \tilde{W}.  
 \end{equation}
Expressions for $S_{NP}$ and $S_{P}$ are in appendix~\ref{app:LL}, in Eqs.~(\ref{eq:cssSudakovNP}) and (\ref{eq:cssSudakovPFull}).
For partons $i,k$, $\tilde{W}$ is
\begin{eqnarray}
 \tilde{W} = \sum_{i,k=q,\overline{q}',g}\left[\mathcal{F}^{T}_{i/p}(\xi_1,b^2,\mu_f^2) \mathcal{F}^{T}_{k/p}(\xi_2,b^2,\mu_f^2) 
 + (\xi_1 \leftrightarrow \xi_2)\right].
\end{eqnarray}
$\mathcal{F}^T$ are the Fourier transformed transverse momentum-dependent (TMD) PDFs evolved to
impact scale $b$ and collinear factorization scale $\mu_f$.
To LL, $\tilde{W}$ is given in Eq.~(\ref{eq:cssResumW}).

The resummed result describes well the $q_T\ll\mStar$ behavior due to the Sudakov suppression.
But because of neglected terms proportional to powers of $q_T$, it underestimates the spectrum at $q_T\gtrsim\mStar$, 
precisely where the FO calculation becomes reliable. 
To describe accurately $q_T$ everywhere, 
one introduces the auxiliary function $d\sigma^{\rm Asymp}$ that matches the asymptotic FO (resummed) behavior at small (large) $q_T$.
Combining the three expressions, the total, matched spectrum is given by~\cite{Arnold:1990yk,Han:1991sa}
\begin{eqnarray}
  \frac{d\sigma^{\rm Matched}}{ 	dq_T} &=&  
      \frac{d\sigma^{\rm Resum}}{	dq_T} 
    + \frac{d\sigma^{\rm FO}}{		dq_T} 
    - \frac{d\sigma^{\rm Asymp}}{	dq_T}.\label{eq:cssCombo}
\end{eqnarray}
The area bound by the $d\sigma^{\rm Matched}$ curve is then normalized to the total $\sigma^{NLO}$ rate of Eq.~(\ref{eq:SigmaNLO})~\cite{Dreiner:2006sv}.
Individual terms of Eq.~(\ref{eq:cssCombo}) are given in Eqs.~(\ref{eq:cssMatchInt}),~(\ref{eq:cssLL}), and ~(\ref{eq:cssAsymp}).

\section{Results}\label{sec:results}
Tree-level results are calculated using helicity amplitudes.
The Cuba library~\cite{Hahn:2004fe} is used for Monte Carlo integration; numerical uncertainty is negligibly small.
Events are output in Les Houches Event (LHE) format~\cite{Alwall:2006yp}.
Rates and shapes are checked by implementing the Lagrangian of Eq.~(\ref{eq:SeesawLag}) 
into FeynRules 2.0.6~\cite{Alloul:2013bka,Christensen:2008py} and using MadGraph$\_$aMC$@$NLO v5.2.1.0~\cite{Alwall:2014hca} (MG5).
Rates are also in agreement with literature~\cite{Arhrib:2009mz}.
We take as SM inputs~\cite{Beringer:1900zz}
\begin{eqnarray}
 M_{Z} = 91.1876\GeV, \quad
\alpha^{\rm \overline{MS}}(M_{Z}) = 1/127.944, \quad
\sin^{2}_{\rm \overline{MS}}(\theta_{W}) = 0.23116.
\end{eqnarray}
The Cabbibo-Kobayashi-Masakawa (CKM) matrix is taken to be unity,
introducing a percent-level error that is no larger than the estimated $\mathcal{O}(\alpha_s^2)$ contributions.
The CJ12mid NLO parton distribution functions (PDFs)~\cite{Owens:2012bv} with $\alpha_s(M_Z)=0.118$ are used. 
Since $2m_T>M_W,M_Z$, the factorization $(\mu_f)$ and renormalization $(\mu_r)$ scales are fixed to the sum of the triplet lepton masses
\begin{equation}
 \mu_0 = \mu_f = \mu_r = m_{T} + m_{\overline{T}} = 2m_{T}.
 \label{eq:renfactScale}
\end{equation}
$\alpha_s(M_Z)$ is run to $\mu_0$ at one-loop in QCD with $n_f=5$.
Setting $n_f = 6$ increases the three-body channel by $2-3\%$, but the total NLO cross section by less than $+1\%$.
For total cross section calculations, we choose soft and collinear cutoffs
\begin{equation}
 \delta_S = 1.0\times10^{-3} \quad\text{and}\quad \delta_C = \delta_S / 100.
\end{equation}
PSS involves fine cancellation of large numbers; for differential distributions, $\delta_S$ is relaxed to 
\begin{equation}
 \delta_S = 3.0\times10^{-3}.
\end{equation}
Born-level $T\overline{T}jX$ events can be generated efficiently by implementing soft and collinear cuts 
Eqs.~(\ref{eq:pssSoftCutDef}) and (\ref{eq:pssColCutDef}) into MG5; see appendix~\ref{app:ThreeBody}. 
Similarly, $T\overline{T}X$ events (sans the hard-collinear PDF subtraction) can be efficiently produced by applying an appropriate scaling. 

For plots in this section, the LO (NLO) curve is denoted by a solid (dashed) line.

\subsection{Total $T^0T^\pm$ and $T^+T^-$ Production at NLO}
\begin{figure}[!t]
\subfigure[]{	\includegraphics[width=0.48\textwidth]{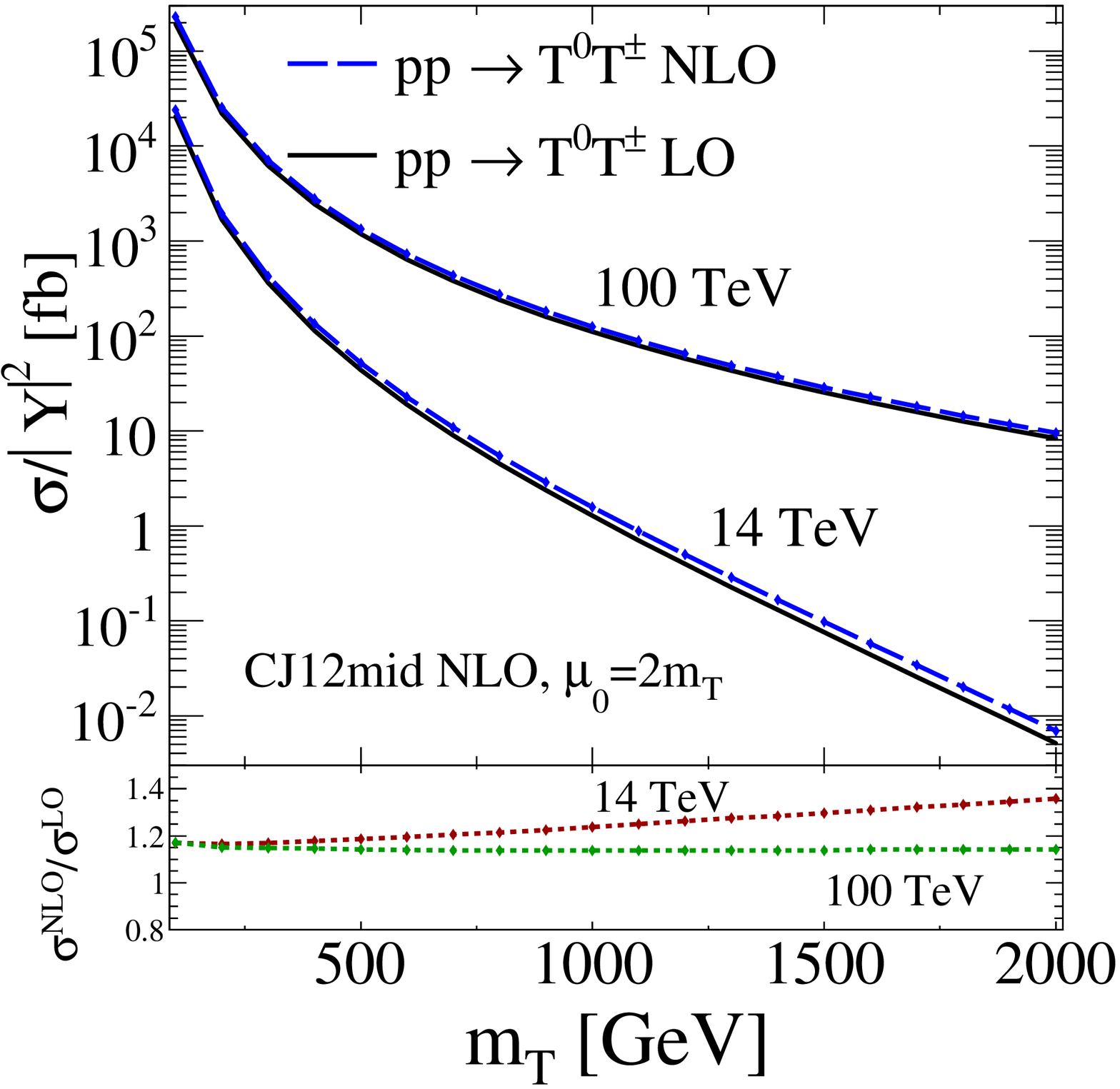}	\label{fig:cc_XSec_vs_Mass}	}
\subfigure[]{	\includegraphics[width=0.48\textwidth]{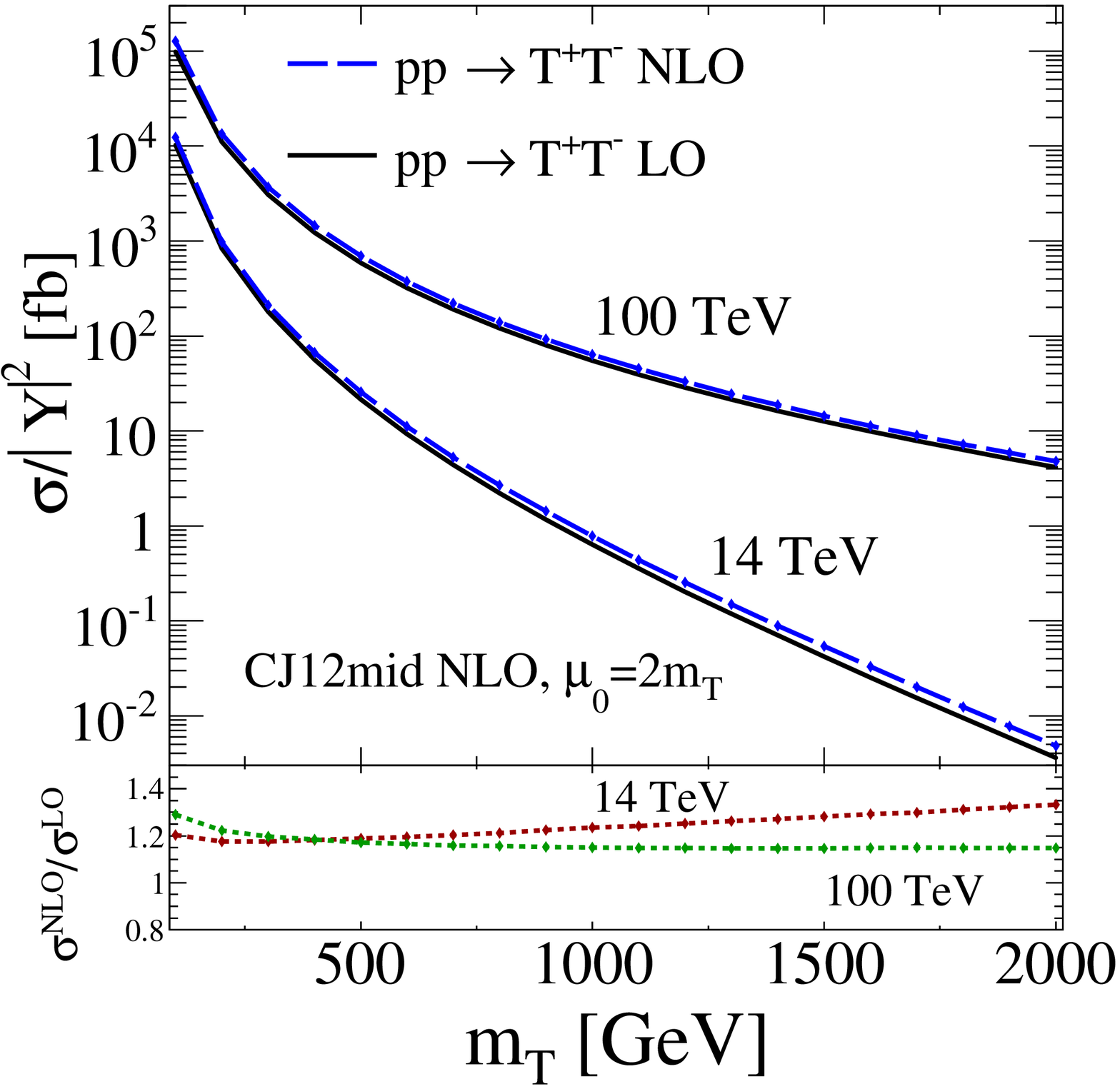}	\label{fig:nc_XSec_vs_Mass}	}
\caption{The 14 and 100 TeV total (a) $T^0T^\pm$ and (b) $T^+T^-$ production cross section, divided by $\vert Y\vert^2$, 
at LO (solid) and NLO (dash) as a function of $m_T$. Panel: Ratio of NLO to LO.}
\label{fig:dyResults}
\end{figure}

Figures~\ref{fig:cc_XSec_vs_Mass} and \ref{fig:nc_XSec_vs_Mass} show, respectively, 
the total CC and NC production cross section, divided by the mixing parameter $\vert Y\vert^2$, as a function of heavy lepton mass 
for $\sqrt{s}=14$ and 100 TeV.
For $m_T = 100\GeV-2\TeV$, the NLO $T^0T^\pm$ production rates range
\begin{eqnarray}
7\ab-24\pb~&\text{at}&	14\TeV
 \\
10\fb-230\pb~	&\text{at}&	100\TeV.
\end{eqnarray}
The corresponding NLO $T^+T^-$ cross sections range
\begin{eqnarray}
5\ab-12\pb~&\text{at}&	14\TeV
 \\
5\fb-130\pb~&\text{at}&	100\TeV.
\end{eqnarray}
The CC rate is approximately twice as large as the NC channel due to $W$ boson charge multiplicity.
In the low-(high-)mass range, transitioning from 14 to 100 TeV 
increases the total cross section roughly by a factor of $10~(1000)$ for both processes.

\begin{table}[!t]
\caption{$T\overline{T}$ NLO Cross Sections and NLO $K$-Factors for Various $pp$ Collider Configurations}
 \begin{center}
\begin{tabular}{|c|c|c|c|c|c|c|}
\hline\hline
\multicolumn{7}{|c|}{$T^0T^\pm$}\tabularnewline\hline
  & \multicolumn{2}{|c|}{13 TeV} & \multicolumn{2}{|c|}{14 TeV} & \multicolumn{2}{|c|}{100 TeV}
\tabularnewline\hline
$m_T$ [GeV]	& $\sigma^{NLO}/\vert Y\vert^2$ [fb] & $K^{NLO}$ & $\sigma^{NLO}/\vert Y\vert^2$ [fb] &
$K^{NLO}$ & $\sigma^{NLO}/\vert Y\vert^2$ [fb] & $K^{NLO}$
	\tabularnewline\hline
100  & $21.7\times10^{3} $  	& $1.17 $    & $23.9\times10^{3} $	& $1.17 $ 	& $229 \times10^{3}$	& $1.17$	\tabularnewline\hline
300  & $370 $  			& $1.17 $    & $423  $			& $1.17 $ 	& $7.07\times10^{3}$	& $1.15$	\tabularnewline\hline
500  & $43.7 $ 			& $1.19 $    & $51.9 $			& $1.19 $ 	& $1.34\times10^{3}$	& $1.14$	\tabularnewline\hline
700  & $8.75 $ 			& $1.21 $    & $10.8 $			& $1.20 $ 	& $433$			& $1.14$	\tabularnewline\hline
900  & $2.22 $ 			& $1.23 $    & $2.89 $			& $1.23 $ 	& $182$			& $1.14$	\tabularnewline\hline
1000 & $1.18 $ 			& $1.25 $    & $1.58 $			& $1.24 $ 	& $125$			& $1.14$	\tabularnewline\hline
1500 & $62.4\times10^{-3} $ 	& $1.31 $    & $97.8\times10^{-3} $	& $1.30 $ 	& $28.8$		& $1.14$	\tabularnewline\hline
2000 & $3.60\times10^{-3} $ 	& $1.37 $    & $6.96\times10^{-3} $	& $1.36 $ 	& $9.58$		& $1.14$	\tabularnewline\hline
\hline
\multicolumn{7}{|c|}{$T^+T^-$}\tabularnewline\hline
  & \multicolumn{2}{|c|}{13 TeV} & \multicolumn{2}{|c|}{14 TeV} & \multicolumn{2}{|c|}{100 TeV}
\tabularnewline\hline
$m_T$ [GeV]	& $\sigma^{NLO}/\vert Y\vert^2$ [fb] & $K^{NLO}$ & $\sigma^{NLO}/\vert Y\vert^2$ [fb] & $K^{NLO}$ & $\sigma^{NLO}/\vert Y\vert^2$ [fb] & $K^{NLO}$
	\tabularnewline\hline
100  & $11.2\times10^{3} $	& $1.20 $ & $12.4\times10^{3} $	& $1.20 $	& $127 \times10^{3}$	& $1.29 $	\tabularnewline\hline
300  & $184 $			& $1.18 $ & $211 $		& $1.18 $	& $3.70\times10^{3}$	& $1.20 $	\tabularnewline\hline
500  & $21.4 $			& $1.19 $ & $25.5 $		& $1.19 $	& $692 $		& $1.17 $	\tabularnewline\hline
700  & $4.26 $			& $1.21 $ & $5.30 $		& $1.20 $	& $221 $		& $1.16 $	\tabularnewline\hline
900  & $1.10 $			& $1.23 $ & $1.42 $		& $1.22 $	& $92.3 $		& $1.15 $	\tabularnewline\hline
1000 & $588 \times10^{-3}$	& $1.24 $ & $782 \times10^{-3} $& $1.23 $	& $63.5 $		& $1.15 $	\tabularnewline\hline
1500 & $35.7\times10^{-3}$	& $1.29 $ & $53.7\times10^{-3}$	& $1.28 $	& $14.4 $		& $1.15 $	\tabularnewline\hline
2000 & $2.74\times10^{-3}$	& $1.35 $ & $4.81\times10^{-3}$	& $1.33 $	& $4.78 $		& $1.15 $	\tabularnewline\hline
\hline
\end{tabular}
\label{tb:kFactor}
\end{center}
\end{table}

The panels of figure~\ref{fig:dyResults} show the NLO $K$-factor at 14 and 100 TeV. For $T^0T^\pm$, the ratios span
\begin{eqnarray}
 1.17 - 1.36 ~&\text{at}&	14\TeV
 \\
 1.14 - 1.17 ~&\text{at}&	100\TeV.
\end{eqnarray}
For $T^+T^-$, they range
\begin{eqnarray}
 1.19 - 1.33 ~&\text{at}&	14\TeV
 \\
 1.15 - 1.29 ~&\text{at}&	100\TeV.
\end{eqnarray}
At lower collider energies, $K$-factors are larger for heavier $m_T$ due to the rarity of antiquarks possessing sufficiently large momentum at LO. 
At NLO, this is compensated by large Bjorken-$x$ gluons undergoing high-$p_T$ $g\rightarrow \overline{q}$ splitting.
CC and NC $K$-factors are appreciable and, due to their color structures, comparable to those of the Seesaw Types I~\cite{Alva:2014gxa}
and II~\cite{Muhlleitner:2003me,Sullivan:2002jt,Jezo:2014wra,Gavin:2010az,Gavin:2012sy}.
Table~\ref{tb:kFactor} summarizes these results for representative $m_T$ at $\sqrt{s}=13$, 14, and 100 TeV $pp$ collider configurations.

\subsection{Scale Dependence}\label{sec:ScaleDependence}
\begin{figure}[!t]
\subfigure[]{	\includegraphics[width=0.48\textwidth]{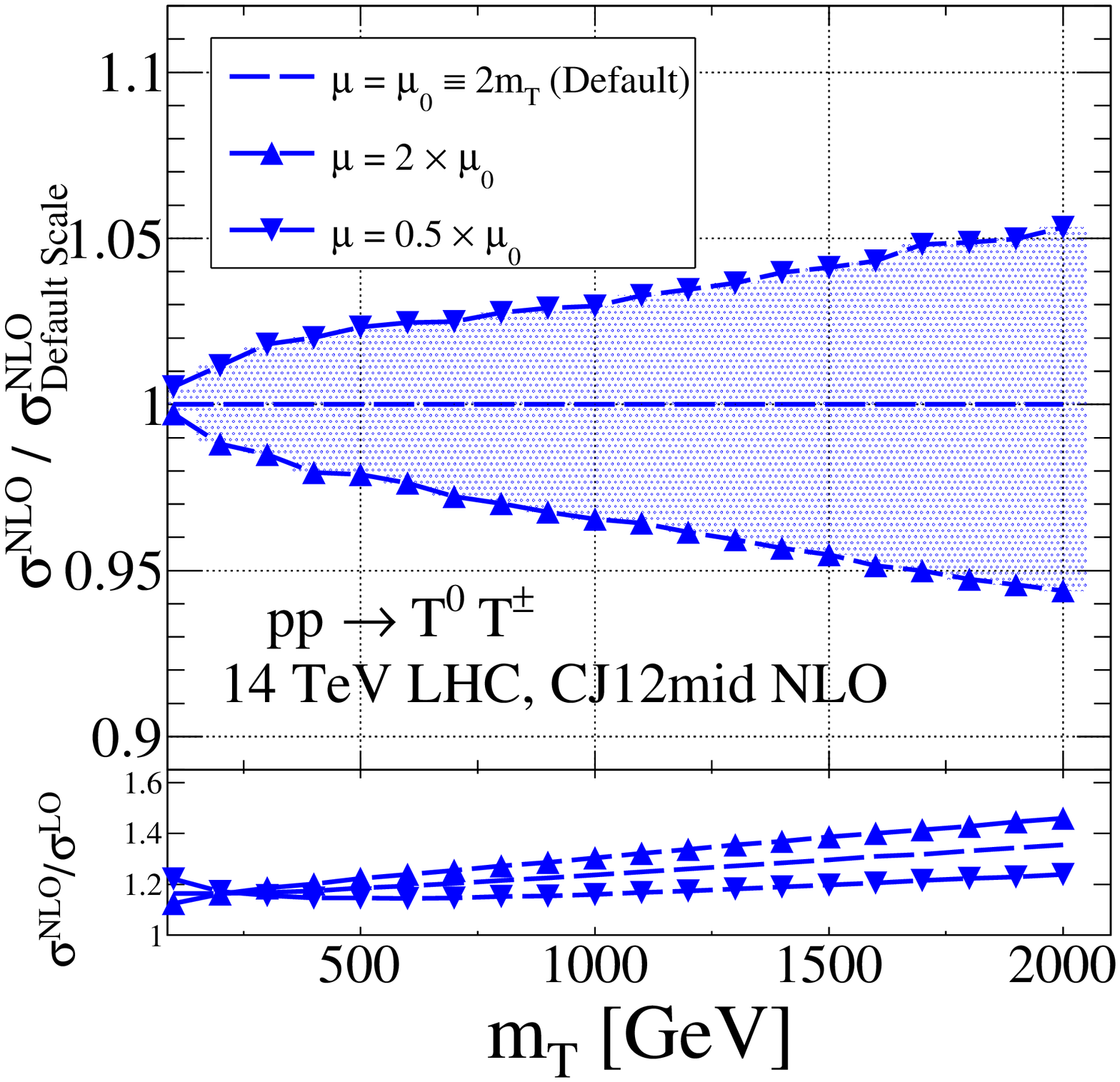}	\label{fig:cc_14TeV_Scale_vs_Mass}	}
\subfigure[]{	\includegraphics[width=0.48\textwidth]{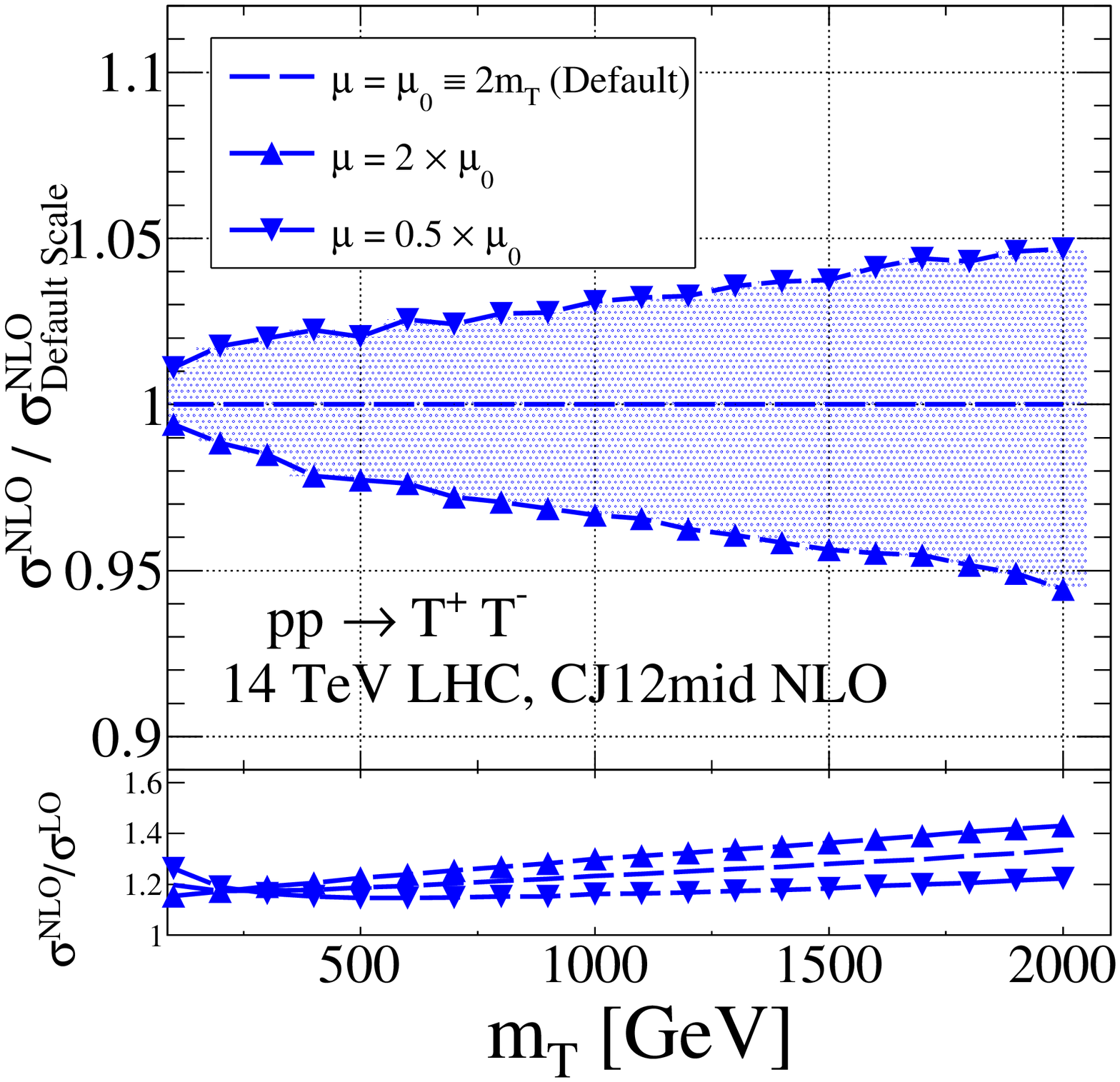}	\label{fig:nc_14TeV_Scale_vs_Mass}	}
\caption{Scale dependence of (a) $T^0T^\pm$ and (b) $T^+T^-$ production at NLO as a function $m_T$.}
\label{fig:scaleResults}
\end{figure}

\begin{table}[!t]
\caption{14 TeV Scale Dependence of $T\overline{T}$ Production Cross Sections at NLO}
 \begin{center}
\begin{tabular}{|c|c|c|c|c|c|c|}
\hline\hline
\multirow{2}{*}{Channel} & \multirow{2}{*}{Scale Choice $\mu$} 	& \multicolumn{5}{|c|}{$m_T$ [GeV]}\tabularnewline\cline{3-7}
 & & $500$ 	& $1000$ & $1500$ & $2000$ 	 	& $2000$ (100 TeV) \tabularnewline\hline\hline
\multirow{2}{*}{$T^0T^\pm$} 	& $0.5\times\mu_0$ 	& $+2.3\%$ & $+3.0\%$ & $+4.1\%$ & $+5.3\%$ 	& $+1.3\%$	\tabularnewline\cline{2-7}
				& $2\times\mu_0$	& $-2.1\%$ & $-3.5\%$ & $-4.5\%$ & $-5.6\%$ 	& $-1.3\%$	\tabularnewline\hline
\multirow{2}{*}{$T^+T^-$} 	& $0.5\times\mu_0$ 	& $+2.0\%$ & $+3.1\%$ & $+3.7\%$ & $+4.7\%$ 	& $+1.4\%$	\tabularnewline\cline{2-7}
				& $2\times\mu_0$	& $-2.3\%$ & $-3.3\%$ & $-4.4\%$ & $-5.6\%$ 	& $-1.2\%$	\tabularnewline\hline
		\hline
\end{tabular}
\label{tb:ScaleDependence}
\end{center}
\end{table}

Higher order QCD corrections are necessary to further reduce theoretical uncertainty.
To quantify and estimate the size of these contributions, the default scale $\mu_0=2m_T$ is varied over the range
\begin{equation}
 \mu_0 \times 0.5 \leq \mu \leq \mu_0 \times 2.
\end{equation}
The scale variation is then defined as the ratio of the NLO rate evaluated at scale $\mu$ to the same rate at $\mu=\mu_0$.
Figures~\ref{fig:cc_14TeV_Scale_vs_Mass} and \ref{fig:nc_14TeV_Scale_vs_Mass} show, respectively, 
the scale variation band of the CC and NC processes at NLO as a function of $m_T$; the lower panels show the NLO $K$-factor for the three scale choices.
The default scale is denoted by a solid line at $1$. 
The high (low) scale scheme is denoted by right-side up (upside down) triangles and is found to decrease (increase) the total cross section.
This suggests that the renormalization scale evolves $\alpha_s(\mu^2)$ to smaller values faster than the factorization scale evolves PDFs to larger values,
and also leads (accidentally) to a vanishing scale dependence for $m_T\sim100\GeV$.
The behavior is consistent with other TeV-scale Seesaws mechanisms~\cite{Alva:2014gxa}.
For the $m_T$ studied, the CC and NC calculations exhibit maximally a $^{+5\%}_{-6\%}$ scale dependence at 14 TeV;
this reduces to $\pm1\%$ at 100 TeV for the same mass range considered.
The 14 TeV NLO $K$-factor varies maximally $^{+7\%}_{-8\%}$ for the CC process and $^{+8\%}_{-8\%}$ for the NC process.
The slightly larger scale dependence in the $K$-factors than the total cross sections is due to the larger scale dependence of the LO result.
The size of the $\mu$-dependence suggests that $\mathcal{O}(\alpha_s^2)$ effects are small, 
consistent with Refs.~\cite{Alva:2014gxa,Gavin:2010az,Gavin:2012sy}.
Results are summarized in table~\ref{tb:ScaleDependence} for representative $m_T$.

\subsection{14 TeV Kinematic Distributions at NLO and NLO+LL}\label{sec:Kinematics}
\begin{figure}[!t]
\subfigure[]{	\includegraphics[width=0.48\textwidth]{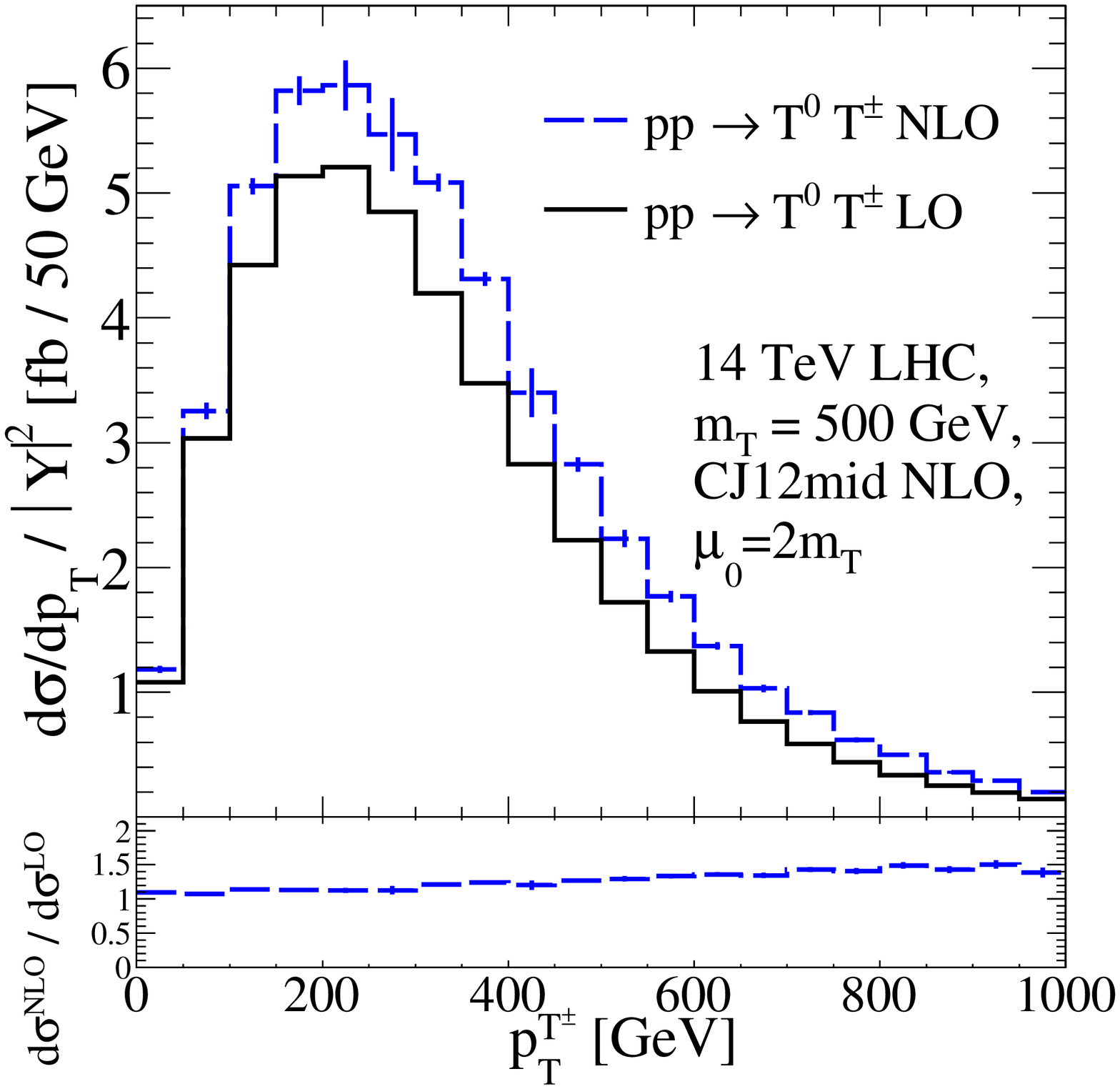}	\label{fig:cc_14TeV_mT500GeV_dpTTpm}	}
\subfigure[]{	\includegraphics[width=0.48\textwidth]{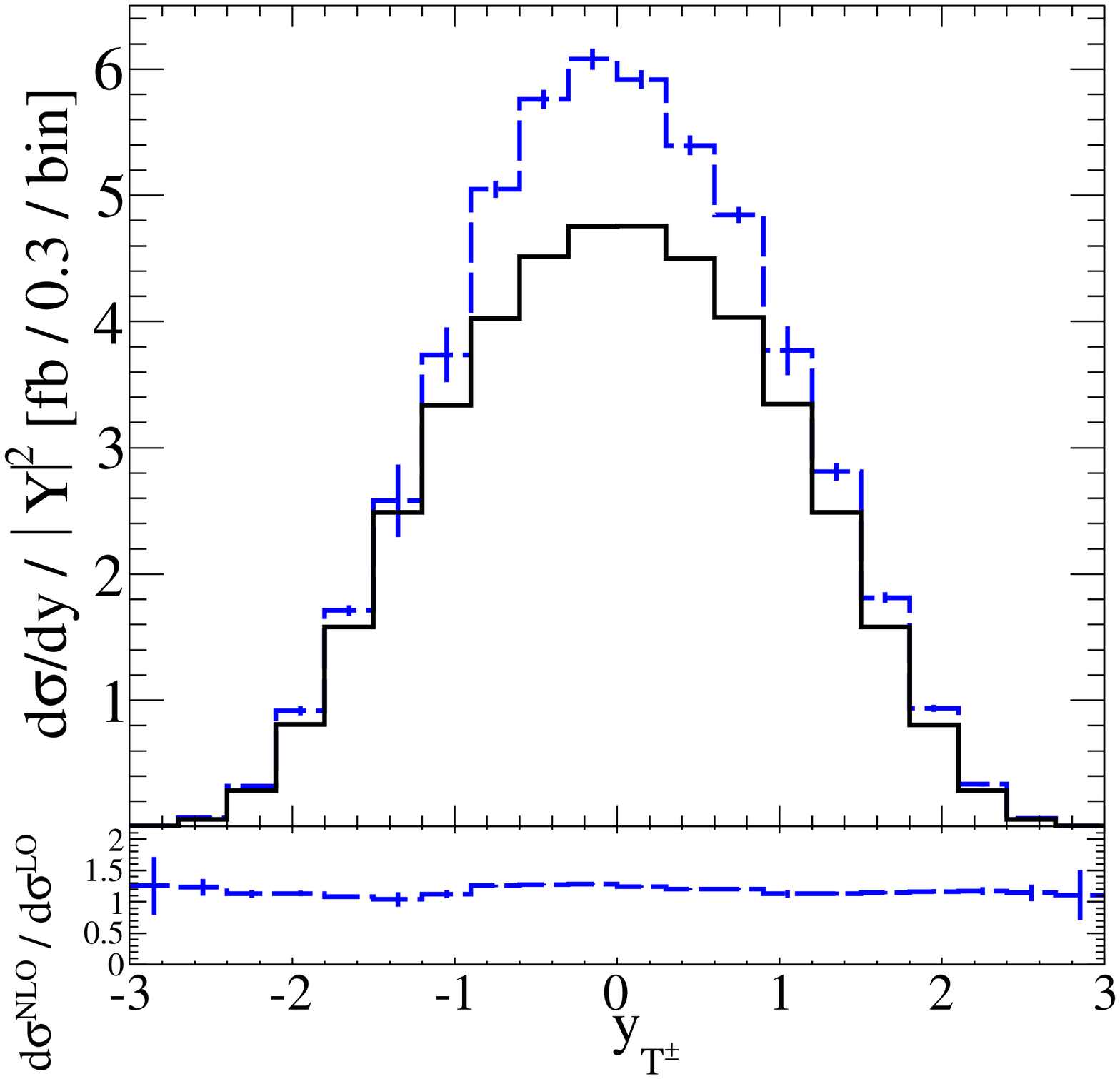}	\label{fig:cc_14TeV_mT500GeV_dyTpm}	}
\caption{
The 14 TeV differential distribution, divided by $\vert Y\vert^2$, with respect to (a) $p_T$ and (b) $y$ of $T^\pm$ in $T^0T^\pm$ production 
at LO (solid) and NLO (dash). Panel: Ratio of NLO to LO.}
\label{fig:Kinematics}
\end{figure}

Figure~\ref{fig:cc_14TeV_mT500GeV_dpTTpm} shows the 14 TeV NLO differential distribution, divided by the mixing parameter $\vert Y\vert^2$, 
with respect to the $p_T$ of $T^\pm$ in $T^0T^\pm$ production.
The panel shows the differential NLO $K$-factors.
At low (high) $p_T$ the change at NLO is small (large)
and follows from the $T^0T^\pm j$ channel at $\mathcal{O}(\alpha_s)$.
The transverse recoil of dilepton system from hard ISR propagates to individual leptons, thereby providing an additional transverse boost.
As the jet energy softens or is radiated more collinearly to its progenitor, its $p_T$ vanishes, and kinematics at NLO approach those at LO.

In figure~\ref{fig:cc_14TeV_mT500GeV_dyTpm}, the rapidity $(y)$ distribution of $T^\pm$ in $T^0T^\pm$ production is presented.
We observe that QCD corrections have a small impact on the rapidity distribution shape as indicated by the mostly flat $K$-factor,
with only a slight upwards bump at small $y$, where $p_T \gg p_z$.
Similar to the $p_T$ spectrum, new kinematic channels at $\mathcal{O}(\alpha_s)$ all involve high-$p_T$ ISR and do not induce longitudinal boosts,
leaving the $y$ distribution shape largely unchanged.

Similar $p_T$ and $y$ behavior are observed for $T^0$ in $T^0T^\pm$ and $T^\pm$ in the NC processes.

\begin{figure}[!t]
\subfigure[]{	\includegraphics[width=0.48\textwidth]{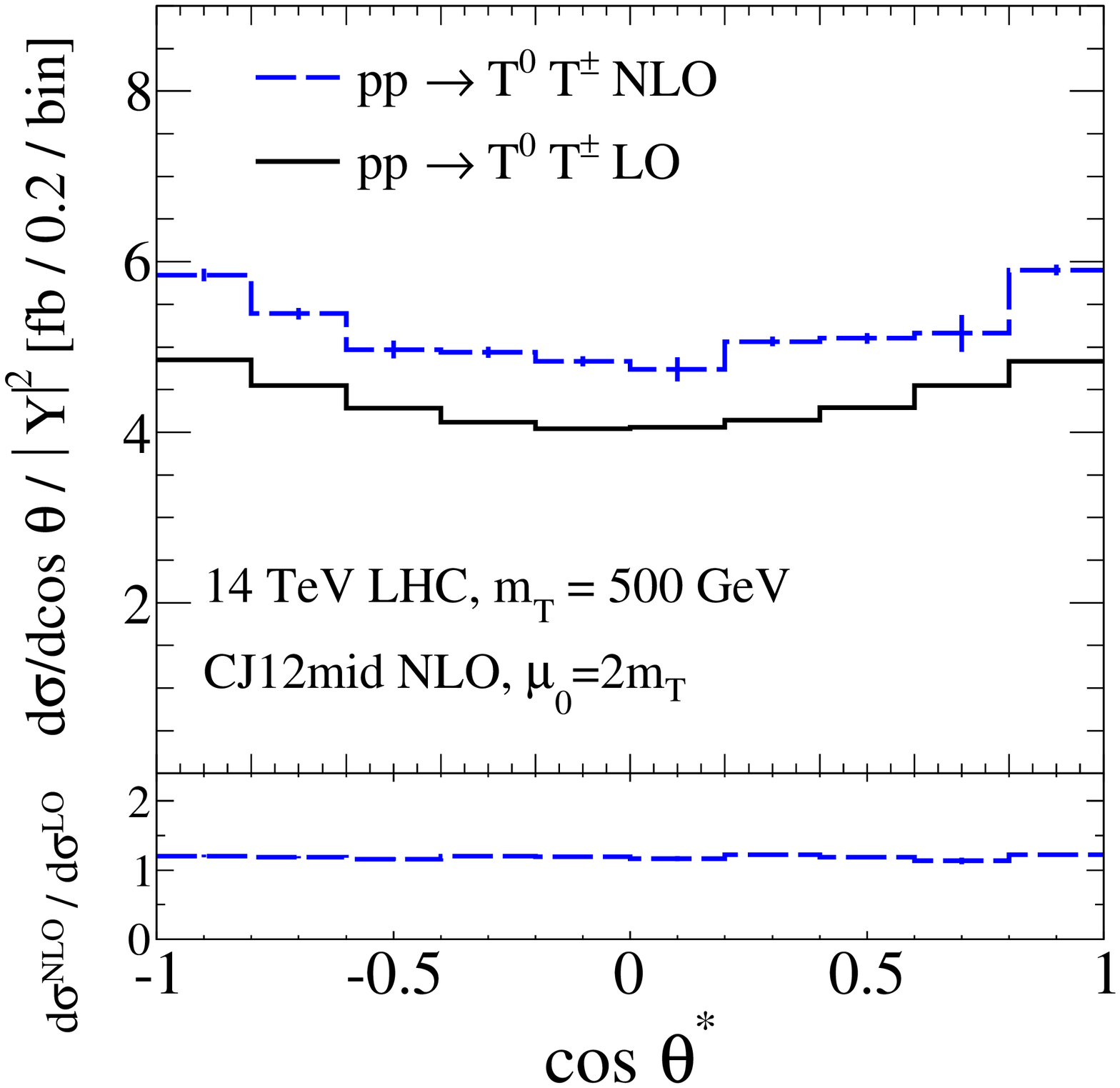}	\label{fig:cc_14TeV_mT500GeV_dcosTh}	}
\subfigure[]{	\includegraphics[width=0.48\textwidth]{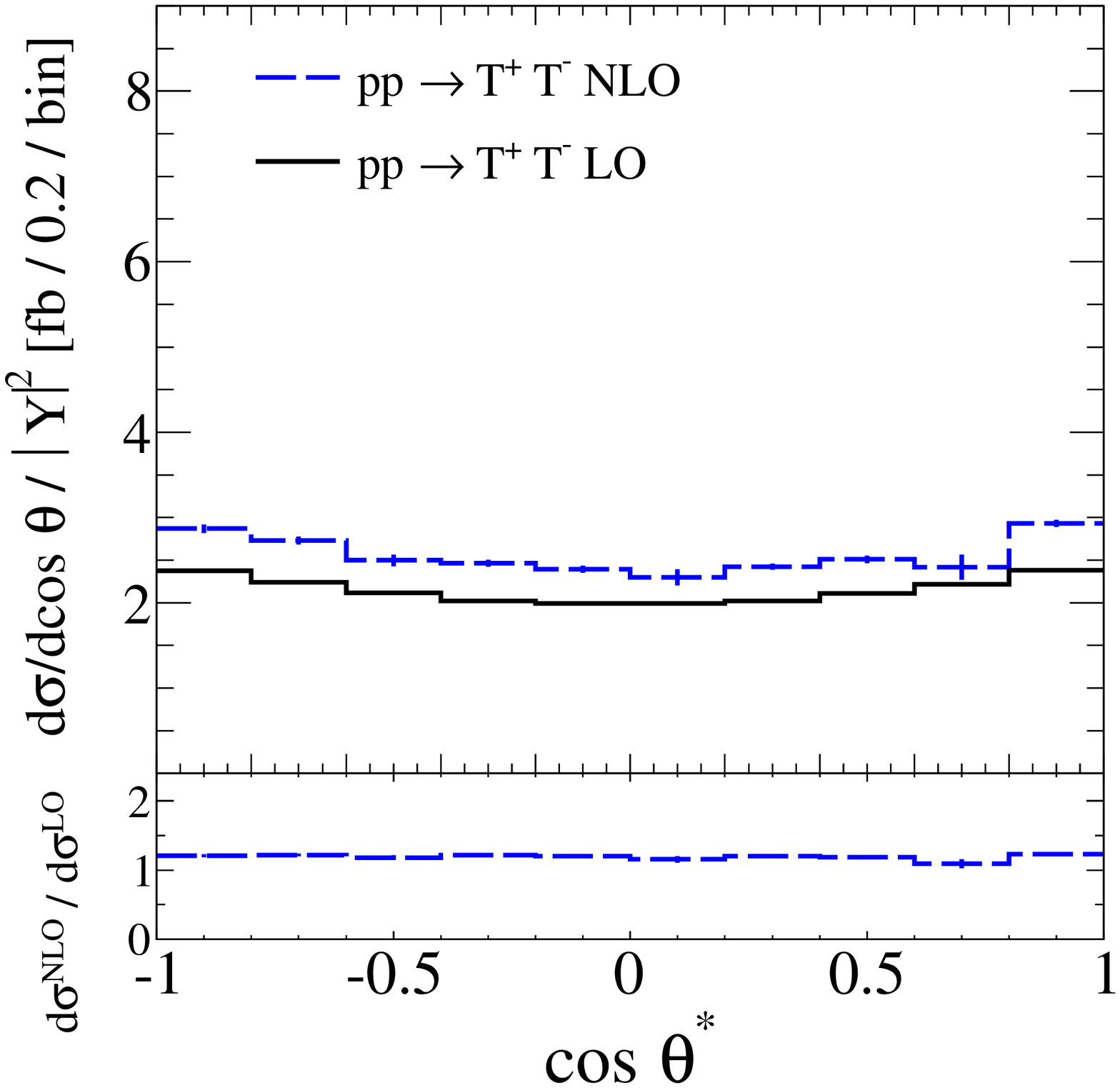}	\label{fig:nc_14TeV_mT500GeV_dcosTh}	}
\caption{
Same as figure~\ref{fig:Kinematics} but with respect to  $\cos\theta^*$, as defined in Eq.~(\ref{eq:PolarDef}), of
(a) $T^\pm$ in $T^0T^\pm$ production and (b) $T^+$ in $T^+T^-$ production.}
\label{fig:AngDist}
\end{figure}

Aside from the Majorana nature of light $(\nu)$ and heavy $(T^0)$ neutral leptons, 
and the relative CC and NC production rates, i.e., $\sigma(pp\rightarrow T^0T^\pm)/\sigma(pp\rightarrow T^+T^-)\sim 2$,
observing vector-like coupling of heavy leptons to electroweak gauge bosons is a critical test of the Type III Seesaw mechanism~\cite{Arhrib:2009mz}.
As in the SM, this done by measuring the polar distribution made by, for example, $T^\pm$ in $T^0T^\pm$ production
in the dilepton rest frame with respect to dilepton system's direction of propagation in the lab frame.
Symbolically, the observable is given by
\begin{equation}
 \cos\theta^* = \frac{\vec{p^*_{T}} \cdot \vec{q}}{\vert\vec{p^*_{T}}\vert \vert\vec{q}\vert},
 \label{eq:PolarDef}
\end{equation}
where $\vec{p^*_{T}}$ is the 3-momentum of lepton $T$ in the $T\overline{T}$ frame and 
$\vec{q} = \vec{p_{T}} + \vec{p_{\overline{T}}}$ in the lab frame.
Figures~\ref{fig:cc_14TeV_mT500GeV_dcosTh} and \ref{fig:nc_14TeV_mT500GeV_dcosTh} show, respectively, the CC and NC $\cos\theta^*$ distribution.
At LO and NLO, the vector coupling structure is clear. 
The uniform $K$-factor follows from $\mathcal{O}(\alpha_s)$ corrections involving only initial-state partons
and amount simply to a boost of the dilepton system in the lab frame.
The $\mathcal{O}(\alpha_s)$ effects are unraveled in constructing Eq.~(\ref{eq:PolarDef}) and subsequently affect only the normalization.

\begin{figure}[!t]
\subfigure[]{	\includegraphics[width=0.48\textwidth]{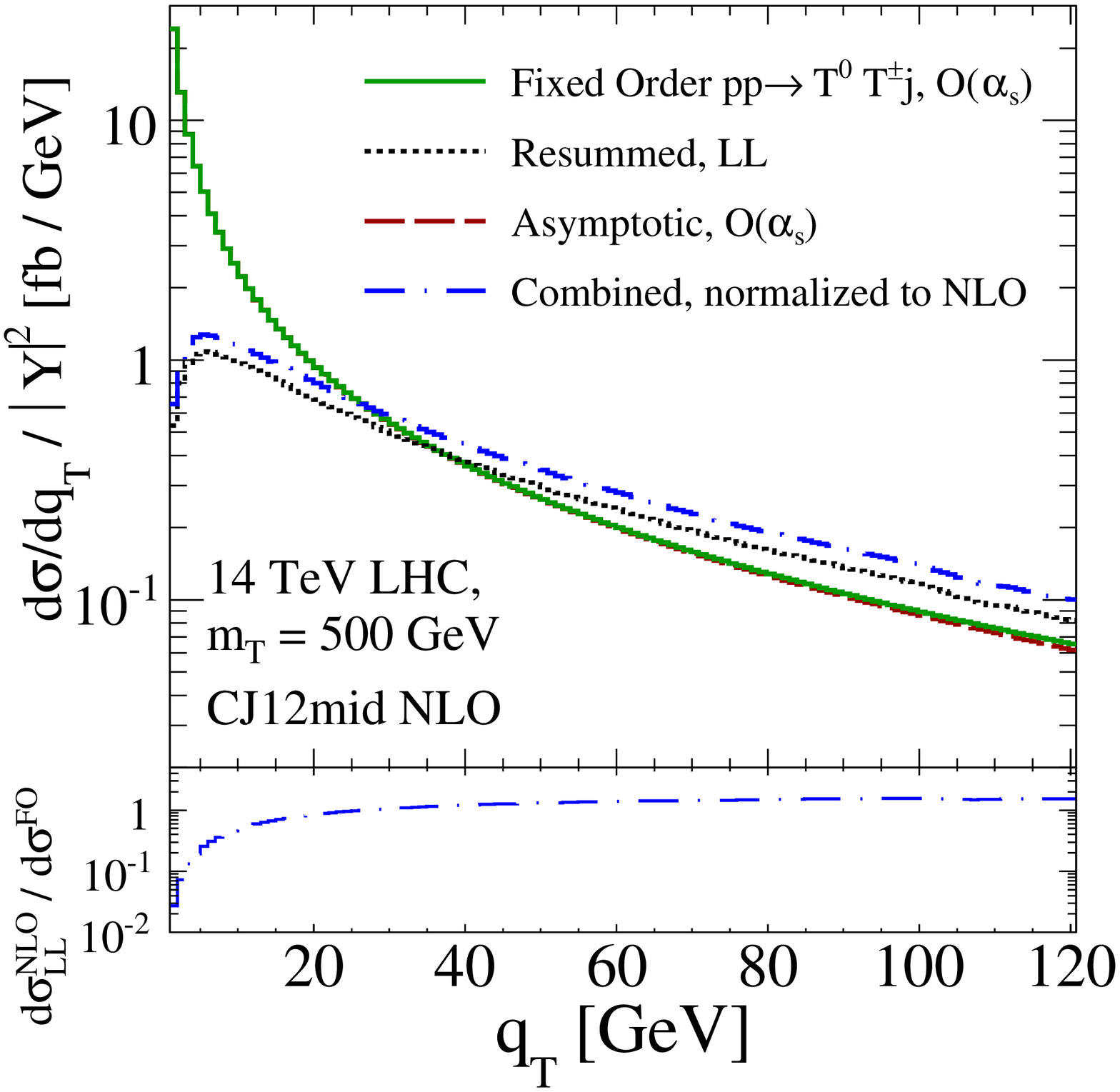}	\label{fig:cc_14TeV_mT500GeV_ResumQT}	}
\subfigure[]{	\includegraphics[width=0.48\textwidth]{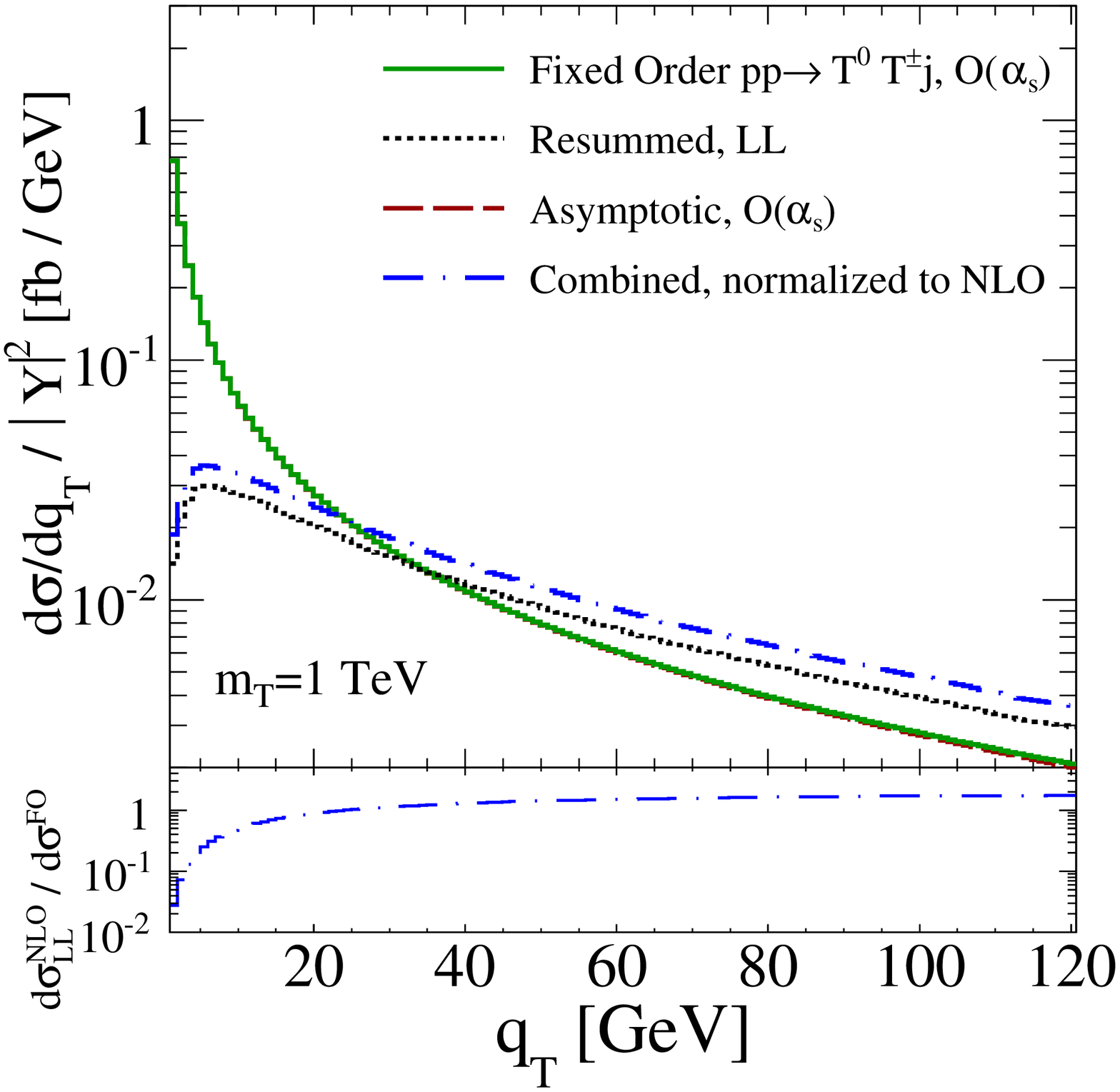}	\label{fig:cc_14TeV_mT1TeV_ResumQT}	}
\caption{
The asymptotic (dash) and FO (solid) at $\mathcal{O}(\alpha_s)$, resummed at LL (dot), 
and combined NLO+LL (dash-dot) contributions to the $T^0T^\pm$ system's transverse momentum differential distribution, $q_T$,
for mass (a) $m_T = 500\GeV$ and (b) $1\TeV$.}
\label{fig:ccLLResults}
\end{figure}

As discussed in section~\ref{sec:cssMethod}, the transverse momentum distribution $q_T$ of the $T^\pm T^0$ dilepton system 
is ill-defined at $q_T\ll\mStar$ in FO perturbation theory and requires resummation of logarithms associated with soft gluon radiation.
For representative masses (a) $m_T = 500\GeV$ and (b) $1\TeV$, figure~\ref{fig:ccLLResults} shows the various contributions to 
$q_T$ spectrum: the asymptotic (dash) and FO (solid) terms at $\mathcal{O}(\alpha_s)$, the resummed rate at LL (dot), and combination of the pieces.
The lower panel shows the ratio of the combined result to the FO result. 
For $m_T=500\GeV~(1\TeV)$, the FO calculation overestimates the combined differential rate for $\confirm{q_T\lesssim 25\GeV}$.
At $q_T\sim 55~(95)\GeV$, the estimate from Eq.~(\ref{eq:ResEstimation}), the FO result remains about \confirm{35\%~(70\%)} below the combined result.
The largeness of the resummation corrections is consistent with other recoil resummations at high-scales~\cite{Dreiner:2006sv}.
The combined distributions peak at $q_T\approx 5-6\GeV$, 
below which the Sudakov suppression from multiple soft gluon emissions overtakes the divergent nature of soft emissions.

\subsection{Discovery Potential at 14 TeV High-Luminosity LHC and 100 TeV}\label{sec:discovery}
In the high luminosity LHC (HL-LHC) scenario~\cite{Brock:2014tja}, detector experiments aim to each collect $1-3\invab$ of data.
We briefly address the maximum sensitivity to heavy triplet lepton pair production in this scenario.
TeV-scale triple leptons decay dominantly to longitudinally polarized weak bosons and the Higgs~\cite{Arhrib:2009mz,Li:2009mw},
a consequence of the Goldstone Equivalence Theorem, implying 
\begin{equation}
 \BR{T^\pm \rightarrow W^\pm \nu_\ell} \approx 2~\BR{T^\pm\rightarrow Z \ell^\pm} 
 \approx 2~\BR{T^\pm\rightarrow h \ell^\pm}\approx 50\%
\end{equation}
\begin{equation}
 \BR{T^0\rightarrow W^+\ell^-+W^-\ell^+} \approx 2~\BR{T^0\rightarrow Z \nu_\ell+Z\overline{\nu}_\ell} 
 \approx 2~\BR{T^0\rightarrow h \nu_\ell+h\overline{\nu}_\ell}\approx 50\%.
\end{equation}
For visible decay modes $Z\rightarrow jj$ and $h\rightarrow b\overline{b}, gg$, heavy lepton pairs can decay into fully reconstructible final-states 
with four jets and two high-$p_T$ leptons that scale like $p_T^\ell \sim m_T/2$:
\begin{eqnarray}
 T^0T^\pm	\rightarrow ~\ell\ell' + WZ/Wh		&\rightarrow& 
 ~\ell\ell' ~+~ 4j ~/~  2j+2b
 \\
 T^+T^- 	\rightarrow ~\ell\ell' + ZZ/Zh/hh 	&\rightarrow& 
 ~\ell\ell' ~+~ 4j ~/~  2j+2b ~/~ 4b.
\end{eqnarray}
The corresponding branching fractions are
\begin{eqnarray}
 &\BR{T^0T^\pm \rightarrow \ell\ell' + 4j/2j+2b} &\approx 11.5\%,
 \\
 &\BR{T^+T^- \rightarrow \ell\ell' + 4j/2j+2b/4b} &\approx 11.6\%.
\end{eqnarray}
Taking $\vert Y\vert^2 = 1$, $\confirm{m_T = 1.5\TeV}$, and acceptance efficiency of $\mathcal{A}=0.75$~\cite{Arhrib:2009mz}, 
then after $3\invab$ one expects tens of heavy lepton pairs across both channels
\begin{eqnarray}
 N_{T^0T^\pm} &=& \mathcal{L} \times \sigma^{NLO}(T^0T^\pm) \times \text{BR} \times \mathcal{A} \approx 25.2,
 \\
 N_{T^+T^-} &=& \mathcal{L} \times \sigma^{NLO}(T^+T^-) \times \text{BR} \times \mathcal{A} \approx 14.0.
\end{eqnarray}
To a good approximation, 
the kinematics of TeV-scale $T\overline{T}$ decays render the SM background negligible~\cite{Arhrib:2009mz,Li:2009mw}.
Using a Gaussian estimator, the statistical significances are at the $3-5\sigma$ level:
\begin{eqnarray}
 \sigma_{T^0T^\pm}(3\invab) &\approx& \sqrt{N_{T^0T^\pm}} = 5.0,
 \\
 \sigma_{T^+T^-}(3\invab) &\approx& \sqrt{N_{T^+T^-}} = 3.7.
\end{eqnarray}
Summing in quadrature, the combined significance surpasses the $6\sigma$ level (over the null hypothesis):
\begin{equation}
 \sigma_{T^0T^\pm + T^+T^-}[3\invab] = \sigma_{T^0T^\pm}[3\invab] \oplus \sigma_{T^+T^-}[3\invab] = 6.3.
\end{equation}
For $\confirm{m_T = 1.6\TeV}$, the combined significance is approximately $4.8\sigma$,
demonstrating a maximum HL-LHC discovery potential to Type III Seesaw leptons in the $m_{T}=1.5-1.6\TeV$ range.

\begin{figure}[!t]
\subfigure[]{	\includegraphics[width=0.48\textwidth]{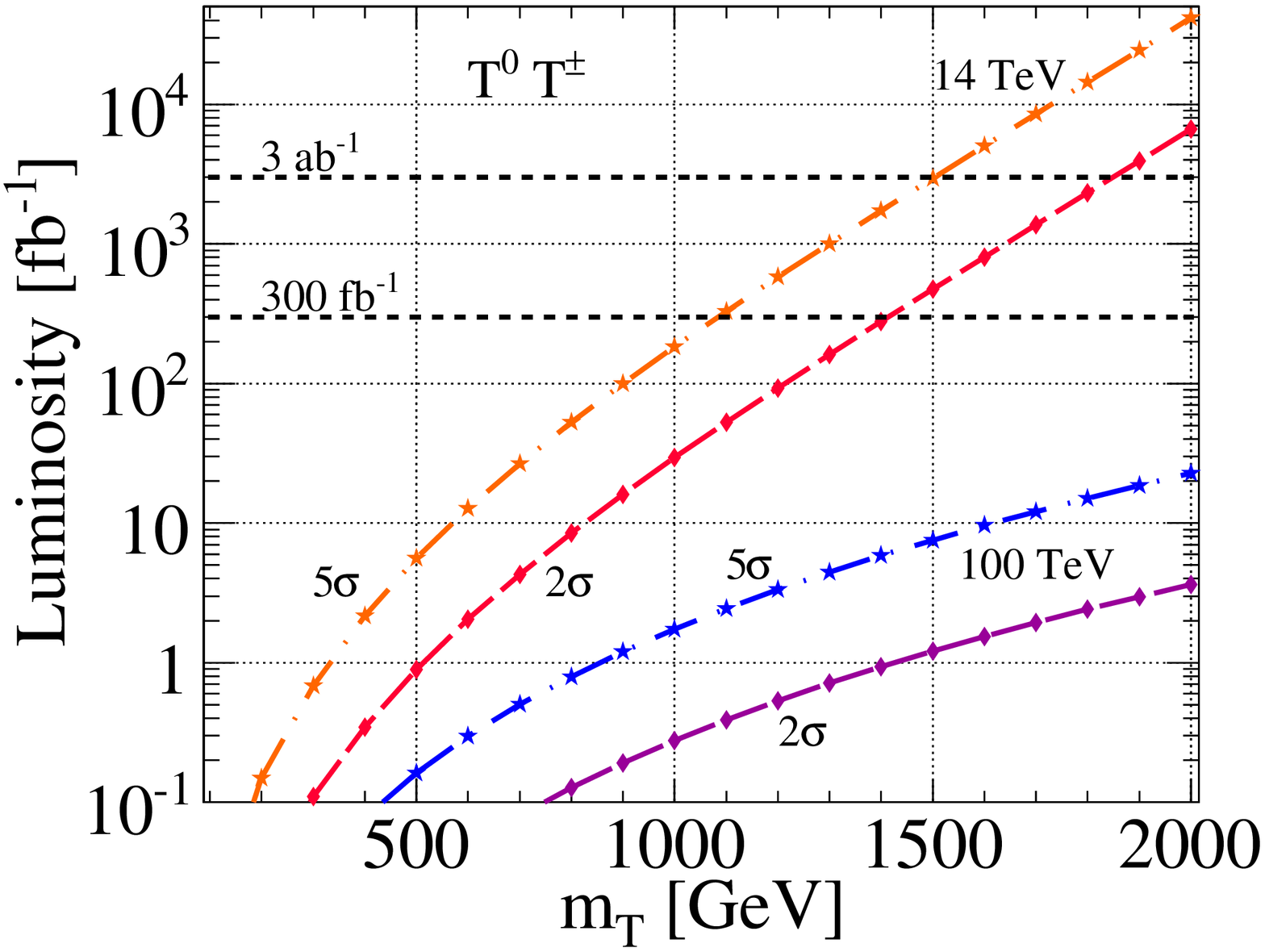}	\label{fig:seesawIII_CC_DiscoveryPotential}	}
\subfigure[]{	\includegraphics[width=0.48\textwidth]{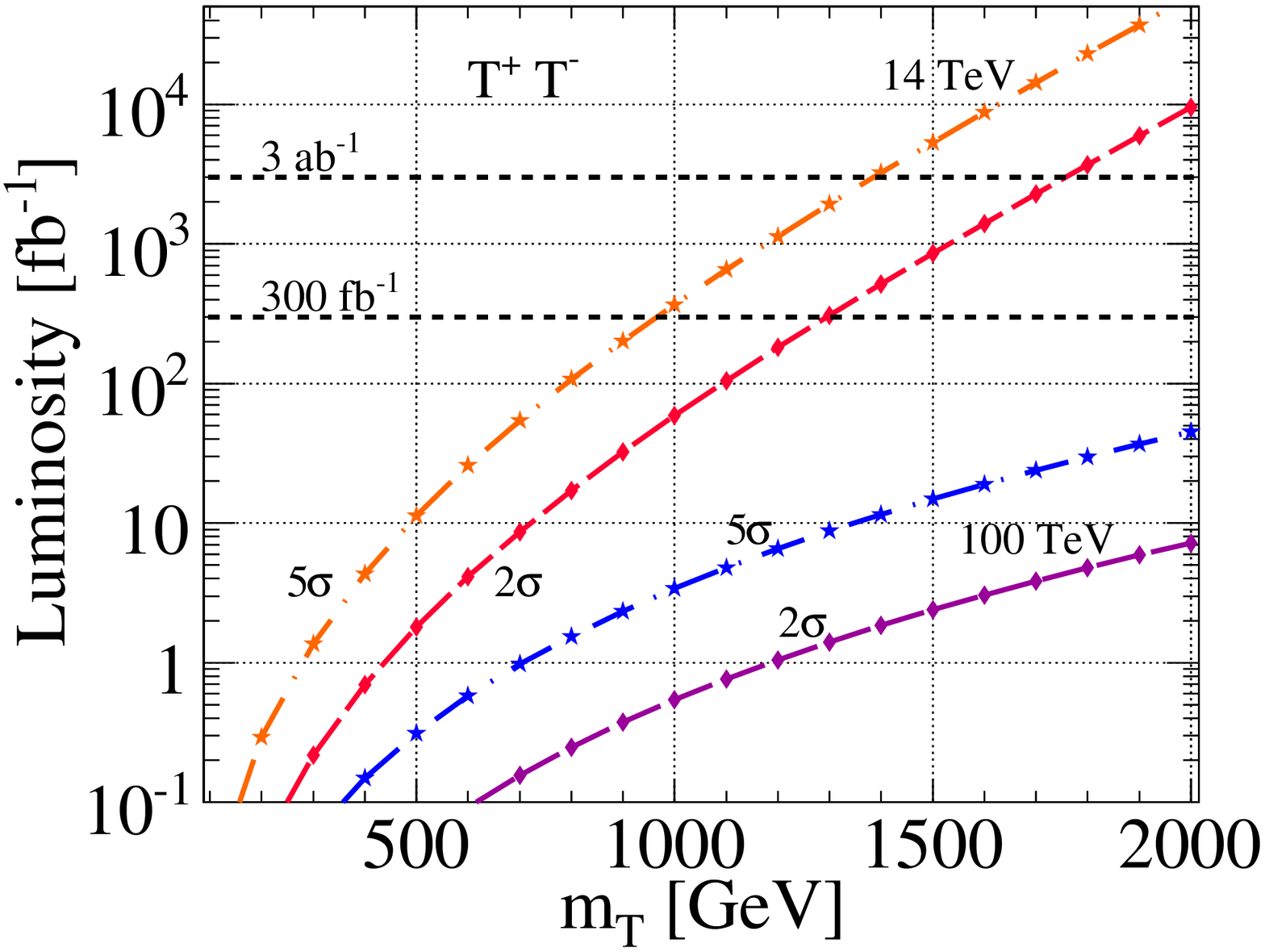}	\label{fig:seesawIII_NC_DiscoveryPotential}	}
\caption{Required luminosity as a function of $m_T$ at 14 and 100 TeV for $5\sigma$ discovery (dash-star) and $2\sigma$ sensitivity (dash-diamond) of 
(a) $T^0T^\pm$ and (b) $T^+T^-$ production, assuming branching fractions and acceptance given in section~\ref{sec:discovery}.}
\label{fig:discovery}
\end{figure}

Fixing the branching fractions and acceptance rates, we plot in figure~\ref{fig:discovery} the 
required luminosity as a function of $m_T$ for a $5\sigma$ discovery (dash-star) and $2\sigma$ sensitivity (dash-diamond) of 
the (a) $T^0T^\pm$ and (b) $T^+T^-$ channels at 14 and 100 TeV.
At 14 TeV and after 300 (3000)$\invfb$, one finds sensitivity to triplet pairs up to $m_T = 1.3-1.4\TeV~(1.7-1.8\TeV)$ in the individual CC and NC channels.
At 100 TeV, we observe that with 10$\invfb$ a $5\sigma$ discovery can be achieved in the CC channel for $m_T\approx1.6\TeV$ 
and in the NC channel for $m_T\approx1.4\TeV$.
At large $m_T$, however, taking $\mathcal{A}=0.75$ is a not justified 
as a different search methodology is required to account for the boosted kinematics of the $T\overline{T}$ decay products.

\section{Summary and Conclusions}\label{sec:summary}
The existence of tiny neutrino masses has broad impact in cosmology, particle, and nuclear physics.
Hence, understanding the origin of their sub-eV masses is a pressing issue.
We report the leading QCD corrections to the production rate and kinematic distributions of 
hypothetical, heavy Type III Seesaw lepton pairs, including soft gluon resummation of the heavy dilepton system. 
We find:
\begin{enumerate}
 \item The $pp\rightarrow T^0T^\pm$ and $T^+T^-$ $K$-factors at NLO in QCD (see table~\ref{tb:kFactor}) span
 \begin{eqnarray}
  T^0T^\pm 	&:& 1.17-1.37~(1.17-1.36)~[1.14-1.17]\quad\text{at}\quad\sqrt{s} = 13~(14)~[100]\TeV,
  \\
  T^+T^- 	&:& 1.18-1.35~(1.19-1.33)~[1.15-1.29]\quad\text{at}\quad\sqrt{s} = 13~(14)~[100]\TeV.
 \end{eqnarray}

 \item For the range of $m_T$ considered, the total $T^0T^\pm$ and $T^+T^-$ production cross sections 
 exhibit a $^{+5\%}_{-6\%}$ scale dependence at 14 TeV and $\pm1\%$ dependence at 100 TeV; see section~\ref{sec:ScaleDependence}.

 \item Differential $K$-factors with respect to rapidity, transverse momentum, and polar distributions of heavy triplet leptons are largely flat, 
  indicating that na\"ive rescaling of Born-level result is a largely justified estimate of kinematics at $\mathcal{O}(\alpha_s)$; see section~\ref{sec:Kinematics}.

\item The resummed transverse momentum of the $T^\pm T^0$ dilepton system illustrates that the impact of TeV-scale systems recoiling off soft radiation is large; 
see figure~\ref{fig:ccLLResults}.
  
\item $T\overline{T}$ production and decay into the $\ell\ell'+4j$ final-state  has been estimated at the HL-LHC,
showing maximally a $4.8-6.3\sigma$ discovery potential for $m_T = 1.5-1.6\TeV$.
At 14 TeV and after 300 (3000)$\invfb$, there is $2\sigma$ sensitivity up to $m_T = 1.3-1.4\TeV~(1.7-1.8\TeV)$ in the individual CC and NC channels.
At 100 TeV and after 10$\invfb$, a $5\sigma$ discovery can be achieved for $m_T\approx1.4-1.6\TeV$. See section~\ref{sec:discovery}. 

\end{enumerate}

\vspace*{0.3cm}
\noindent {\it Acknowledgments:}
{Shou-Shan Bao, Ben Clark, Ayres Freitas, Tao Han, Josh Isaacson, Josh Sayre, Zhuoni Qian, Zong-Guo Si, and Xing Wang 
are thanked for valuable discussions.
Tao Han and Josh Sayre are thanked for careful readings of the manuscript.
R.R.~acknowledges support from the University of Pittsburgh.}

\hrulefill

\begin{appendix}\label{app:QCD}

\section{Triplet Lepton Pair Production at NLO in QCD via PSS}\label{app:NLO}
The PSS technique and explicit examples, including the DY process, are reviewed authoritatively in Ref.~\cite{Harris:2001sx}.
Terms relevant to triplet lepton pair production in hadron collisions are collected here for convenience.
We emphasize that the NLO calculation greatly simplifies to simply convolving PDFs with tree-level amplitudes.

\subsection{QCD-Corrected Two-Body Final State}
In the soft/collinear limits, radiation in $T\overline{T}j$ production becomes unresolvable and the three-body kinematics approach those of 
the two-body process
\begin{equation}
 q(p_A) ~\overline{q}'(p_B) ~\rightarrow  T(p_1) ~\overline{T}(p_2).
\end{equation}
Subsequently, the singular propagators in the amplitude factor and 
the soft/collinear contributions to the $T\overline{T}j$ cross section can be expressed as a divergent, but process-independent, 
piece and the LO $T\overline{T}$ cross section.
Loop corrections cancel soft and soft-collinear poles rendering finite the quantity
\begin{equation}
 d\hat{\sigma}_{(2)} = d\hat{\sigma}^{B} + d\hat{\sigma}^{V} + d\hat{\sigma}^{S}  + d\hat{\sigma}^{SC}.
\end{equation}
PDFs are redefined to subtract residual hard-collinear divergence. We now give each term explicitly.

\subsubsection{Virtual Corrections}
For massless fermions, $\mathcal{O}(\alpha_s)$ corrections to EW vertices, i.e., figure~\ref{fig:feynman}~(b), 
factorize~\cite{Altarelli:1979ub} into a product of the Born matrix element, $\mathcal{M}_{B}$, and a universal form factor:
\begin{eqnarray}
 \overline{v}(p_d) \gamma^\mu \left(g_L P_L + g_R P_R \right) u(p_u) &\overset{\rm 1-Loop}{\longrightarrow}& \overline{v}(p_d)\Gamma^\mu(p_u,p_d)u(p_u),
 \end{eqnarray}
\begin{eqnarray}
 \overline{v}(p_d)\Gamma^\mu(p_u,p_d)u(p_u) &=& \overline{v}(p_d)\gamma^\mu \left(g_L P_L + g_R P_R \right) u(p_u) \times \mathcal{F}\\
 \mathcal{F} &\equiv& 
 \frac{\alpha_s(\mu_r^2)}{4\pi}C_F C_\vareps(\hat{s})(-1)^\vareps \Gam{1+\vareps}\Gamma\left(1-\vareps\right)
 \left(\frac{-2}{\vareps^2}-\frac{3}{\vareps}-8 \right),
 \\
 C_\vareps(\hat{s}) &=& \left(\frac{4\pi \mu_r^2}{\hat{s}}\right)^\vareps \frac{\Gamma\left(1-\vareps\right)}{\Gam{1-2\vareps}},
 \quad C_F = 4/3.
\end{eqnarray}
The vertex correction is UV-finite, so its counter term is zero.
Counter terms cancel external quark self-energy corrections in the on-shell and $\overline{\rm MS}$ schemes.
Hence, the summed, squared amplitude is
\begin{eqnarray}
 \sum \vert\mathcal{M}\vert^2 &=& \sum\vert\mathcal{M}_{B}\vert^2 + 2\sum\Re\left[\mathcal{M}_{B}^*\mathcal{M}_{\rm 1-Loop}\right] + \mathcal{O}(\alpha_s^2),
 \quad \mathcal{M}_{\rm 1-Loop} = \mathcal{M}_{B}\mathcal{F}
  \\
 &=& \sum\vert\mathcal{M}_{B}\vert^2\left(1+ 2\Re[\mathcal{F}] \right)+ \mathcal{O}(\alpha_s^2).
\end{eqnarray}
The Born and virtual terms at $\mathcal{O}(\alpha_s)$ for heavy lepton production are then
\begin{eqnarray}
  d\hat{\sigma}^B+d\hat{\sigma}^V &=& d\hat{\sigma}^{B}
 \left[1 + \frac{\alpha_s(\mu_r^2)}{2\pi}C_FC_\vareps(\hat{s})
 \left(\frac{A_2^V}{\vareps^2}+\frac{A_1^V}{\vareps} + A_0^V\right)
 \right],
 \\
 A_2^V &=& -2, \quad A_1^V = -3, \quad A_0^V = -8 +\frac{2\pi^2}{3}.
 \label{eq:qdcV}
\end{eqnarray}

\subsubsection{Soft Corrections}
For colored partons $a,b$, and color-connected Born cross section $\hat{\sigma}^0_{ab}$, the soft radiation expression is
\begin{eqnarray}
 d\hat{\sigma}^{S} &=&  \frac{\alpha_s(\mu_r^2)}{2\pi} C_\vareps(\hat{s}) 
 ~\sum_{a,b=q,\overline{q}'} ~d\hat{\sigma}^0_{ab} \int dS \frac{-(p_a\cdot p_b)}{(p_a\cdot p_j)(p_a\cdot p_j)},
\\
&=& (-1)
 \frac{\alpha_s(\mu_r^2)}{2\pi} C_\vareps(\hat{s})  ~  \int dS ~\left[
 \frac{(p_q \cdot p_{\overline{q}'})}{(p_q\cdot p_j)^2}d\hat{\sigma}^0_{\overline{q}'q}	 + 
 \frac{(p_{\overline{q}'} \cdot p_q)}{(p_{\overline{q}'}\cdot p_j)^2}d\hat{\sigma}^0_{q\overline{q}'} \right],
\\
 dS &=& \frac{1}{\pi}\left(\frac{4}{\hat{s}}\right)
 \int_0^{\delta_S \sqrt{\hat{s}}/2}dE_j E_j^{1-2\vareps} ~
 \int d\theta \sin^{1-2\vareps}\theta ~
 \int d\phi \sin^{-2\vareps}\phi,
\end{eqnarray}
where $dS$ is the soft one-particle phase space in $n=4-2\vareps$ dimensions, 
and we make use of the fact that the only two colored partons in the DY process are the initial-state quark and antiquark.
For the present case, the color-connected and plain Born cross sections are related by
\begin{equation}
 d\hat{\sigma}^0_{q\overline{q}'} = d\hat{\sigma}^0_{\overline{q}'q} = - C_F ~d\hat{\sigma}^{B}.
\end{equation}
We take $p_q~(p_{\overline{q}'})$ to propagate in the $+\hat{z}$ ($-\hat{z}$) direction, implying
\begin{eqnarray}
 p_q \cdot p_j &=& \frac{\sqrt{\hat{s}}}{2}E_j (1-\cos\theta), 
 \quad p_{\overline{q}'} \cdot p_j = \frac{\sqrt{\hat{s}}}{2}E_j (1+\cos\theta),
 \quad p_q \cdot p_{\overline{q}'} = \frac{\hat{s}}{2}.
\end{eqnarray}
Using the tabled integrals of Ref.~\cite{Harris:2001sx}, the soft contribution to DY processes is
\begin{eqnarray}
 d\hat{\sigma}^{S} &=& 
 d\hat{\sigma}^{B} \left[\frac{\alpha_s(\mu_r^2)}{2\pi}C_F C_\vareps(\hat{s})
 \left(\frac{A_2^S}{\vareps^2}+\frac{A_1^S}{\vareps} + A_0^S\right)
 \right],
 \\
 A_2^S &=& 2, \quad A_1^S = -4\log\delta_S, \quad A_0^S = 4\log^2\delta_S.
 \label{eq:qdcS}
\end{eqnarray}

\subsubsection{Soft Collinear Corrections}
For $q\overline{q}'$ initial-states, the soft-collinear correction to the Born amplitude is
\begin{eqnarray}
 d\hat{\sigma}^{SC} &=& d\hat{\sigma}^{B} ~\frac{\alpha_s(\mu_r^2)}{2\pi}C_F C_\vareps(\hat{s})
  \nonumber\\
 &\times&
 \left[
 \left(\frac{A_1^{SC}(q\rightarrow qg)}{\vareps} + A_0^{SC}(q\rightarrow qg)\right)
 +
 \left(\frac{A_1^{SC}(\overline{q}\rightarrow \overline{q}g)}{\vareps} + A_0^{SC}(\overline{q}\rightarrow \overline{q}g)\right)
 \right]
 \\
 &=& d\hat{\sigma}^{B} \left[\frac{\alpha_s(\mu_r^2)}{2\pi}C_F C_\vareps(\hat{s})
 \left(\frac{2A_1^{SC}(q\rightarrow qg)}{\vareps} + 2A_0^{SC}(q\rightarrow qg)\right)
 \right],
 \\
 A_1^{SC} &=& \left(2\log\delta_S + \frac{3}{2}\right), \quad  A_0^{SC} = \left(2\log\delta_S + \frac{3}{2}\right)\log\left(\frac{\hat{s}}{\mu^2_f}\right).
 \label{eq:qdcSC}
\end{eqnarray}
The soft-collinear splitting functions, $A(1\rightarrow 2~3)$,
are derived by integrating over the Altarelli-Parisi (AP) splitting functions.
The $q\rightarrow q$ and $\overline{q}'\rightarrow \overline{q}'$ functions are equal by CP-invariance
and therefore are combined in the last step.

\subsubsection{Assembly of Two-Body Corrections}
A finite result emerges after summing the virtual, soft, and soft-collinear corrections,
\begin{eqnarray}
 d\hat{\sigma}^{V} + d\hat{\sigma}^{S}  + d\hat{\sigma}^{SC} &=& d\hat{\sigma}^{B}~\frac{\alpha_s(\mu_r^2)}{2\pi}C_F C_\vareps(\hat{s})
 ~\Bigg[ \left(\frac{A_2^V}{\vareps^2}+\frac{A_2^S}{\vareps^2} \right) 
 + \left(\frac{A_1^V}{\vareps}+\frac{A_1^S}{\vareps} + \frac{2A_1^{SC}(q\rightarrow qg)}{\vareps}\right)
 \nonumber\\
  & & 
  + \left( A_0^V + A_0^S + 2A_0^{SC}(q\rightarrow qg)\right)
 \Bigg],
\end{eqnarray}
and using Eqs.~(\ref{eq:qdcV}),~(\ref{eq:qdcS}), and~(\ref{eq:qdcSC}) to show
\begin{eqnarray}
 A_2^V + A_2^S = 0, \quad A_1^V + A_1^S + A_1^{SC} = 0.
\end{eqnarray}
With poles removed, we send $\vareps\rightarrow0$. The non-hard-collinear, two-body contribution of Eq.~(\ref{eq:pss2BNonHC}) is
\begin{eqnarray}
d\hat{\sigma}_{(2)} &=& d\hat{\sigma}^{B} + d\hat{\sigma}^{V} + d\hat{\sigma}^{S}  + d\hat{\sigma}^{SC}
\\
&=& 
d\hat{\sigma}^{B}~\left[1+\frac{\alpha_s(\mu_r^2)}{2\pi}C_F \left[A_0^V + A_0^S + 2A_0^{SC}(q\rightarrow qg)\right]\right].
\end{eqnarray}

\subsection{Hard-Collinear Subtraction}
Remaining hard-collinear ISR poles are associated with PDFs,
which themselves are all orders resummations of hard-collinear splittings.
The divergences are removed by subtracting out from PDFs redundant $\mathcal{O}(\alpha_s)$ splittings. 
The hard-collinear subtraction is given by
\begin{eqnarray}
\sigma^{HC}(pp\rightarrow T ~\overline{T} ~X) &=&  
 \sum_{a,b=q,\overline{q}} \int_{\tau_0}^1 d\xi_1 \int^1_{\tau_0/\xi_1}d\xi_2 \qquad \hat{\sigma}^{B}(ab\rightarrow T ~\overline{T}) 
 \nonumber\\
 &\times&  \left[
 f_{a/p}(\xi_1,\mu^2)\tilde{f}_{b/p}(\xi_2,\mu^2) + \tilde{f}_{a/p}(\xi_1,\mu^2)f_{b/p}(\xi_2,\mu^2) + (1\leftrightarrow2)\right] , 
\end{eqnarray}
where the redefined, $\mathcal{O}(\alpha_s)$-corrected PDFs $\tilde{f}_{a/p}(\xi,\mu^2)$ are
\begin{eqnarray}
 \tilde{f}_{a/p}(\xi,\mu^2) &=&
 \frac{\alpha_s(\mu^2)}{2\pi} \sum_{b=g,q,\overline{q}'} 
 \int^{1-\delta_S \delta_{ba}}_\xi \frac{dz}{z}  f_{b/p}\left(\frac{\xi}{z},\mu^2\right)\tilde{P}_{ab}(z)
 \\
 &=& \frac{\alpha_s(\mu^2)}{2\pi} \left[
 \int^{1-\delta_S}_x \frac{dz}{z}  f_{q/p}\left(\frac{\xi}{z},\mu^2\right)\tilde{P}_{qq}(z) +  
 \int^{1}_x          \frac{dz}{z}  f_{g/p}\left(\frac{\xi}{z},\mu^2\right)\tilde{P}_{qg}(z) \right].
\label{eq:pssSplitPDF}
\end{eqnarray}
The summation is over all partons that give rise to $q,\overline{q}'$ after one radiation, 
$\delta_{ba}$ is the Kronecker $\delta$-function, 
and the modified AP $i\rightarrow j$ splitting functions $\tilde{P}_{ji}(z)$ have the form
\begin{equation}
\tilde{P}_{ji}(z) = P_{ji}(z) \log\left[\frac{(1-z)}{z}\frac{\delta_C\hat{s}}{\mu^2}\right] - P'_{ji}(z).
\end{equation}
In $n=4-2\vareps$ dimensions, $P$ and $P'$ are related by $P_{ji}(z,\vareps) = P_{ji}(z) + \vareps P'_{ji}(z)$, with
\begin{eqnarray}
 P_{qq}(z) = C_F \frac{(1+z^2)}{1-z}, &\quad& P'_{qq} = -C_F (1-z), 
 \nonumber\\
 P_{qg}(z) = \frac{1}{2}\left[z^2 + (1-z)^2 \right], &\quad& P'_{qg}(z) = -z(1-z).
 \label{eq:pssAP}
\end{eqnarray}

\subsection{Efficient Event Generation of Three-Body, Hard, Non-Collinear Radiation}\label{app:ThreeBody}
Imposing soft/collinear cuts in the partonic c.m.~frame on the processes 
\begin{equation}
 q(p_A)\overline{q}'(p_B),~g(p_A)q(p_B),~g(p_A)\overline{q}'(p_B) \rightarrow  T(p_1) ~\overline{T}(p_2) ~j(p_j)
\end{equation}
regulates all diverges.
$T\overline{T}j$ event generation can be handled efficiently by implementing Eqs.~(\ref{eq:pssSoftCutDef})-(\ref{eq:pssColCutDef}) into MG5. 
This entails inserting into the header of the file \texttt{SubProcesses/cuts.f}:
\begin{quote}\texttt{\noindent
 double precision del$\_$s, ~del$\_$c, sHat, ~ejMin, ~tjMin, ~tmptj
}
\end{quote}
and in the file's body:
\begin{quote}\texttt{\noindent
del$\_$s = 3.0d-3\\
del$\_$c = del$\_$s $*$ 1.0d-2\\
sHat  = SumDot(p(0,1),p(0,2),1d0)\\ 
ejMin = del$\_$s $*$ dSqrt( sHat ) /2.0d0\\
tjMin = del$\_$c $*$       sHat\\
\hfill\\
do i=1,nincoming    \\
do j=nincoming+1,nexternal \\
if(is$\_$a$\_$j(j)) then \\
\hfill\\
if( p(0,j).LT.ejMin ) then\\
passcuts=.false.\\
endif\\
\hfill\\
tmptj = dabs(SumDot(p(0,i),p(0,j),-1d0)) \\
if( tmptj.LT.tjMin ) then\\
passcuts=.false.\\
endif\\
\hfill\\
endif\\
enddo\\
enddo
}
\end{quote}
For $T\overline{T}j$, $\hat{s}_{ij}$ cuts are not needed as all $s$-channel poles are regulated by nonzero $m_T$.

\section{Recoil $q_T$ Resummation with Fixed Order Matching}\label{app:LL}
Individual formulae and steps for resumming recoil logarithms at LL accuracy for DY-type processes  
within the CSS formalism~\cite{Collins:1981uk,Collins:1981va,Collins:1984kg} can be found across 
literature~\cite{Collins:1984kg,Davies:1984sp,Hinchliffe:1988ap,Arnold:1990yk,Kauffman:1991jt,Han:1991sa,Landry:1999an,Landry:2002ix,Dreiner:2006sv}
\footnote{We report and clarify a slight but unfortunate typo in Ref.~\cite{Han:1991sa}: the resummed result is to LL accuracy, not NLL.}.
For conciseness and convenience, they are collected here in terms of generic tree-level quantities.

For the processes
\begin{equation}
  q ~\overline{q}' \rightarrow V^{*} \rightarrow T ~\overline{T}, \quad V \in \{\gamma,Z,W\},
  \label{eq:cssDY}
\end{equation}
the resummed and matched $q_T$ distribution of the $T\overline{T}$ system decomposes into the FO, resummed, and asymptotic pieces; see Eq.~(\ref{eq:cssCombo}).
The oscillatory behavior of $J_0(q_T b)$ and the fine numerical cancellation of the resummed and asymptotic terms when $q_T\gtrsim\mStar$
give rise to an unnecessary computational burden for a contribution that is small compared to the FO piece.
In practice~\cite{Hinchliffe:1988ap,Kauffman:1991jt}, 
one introduces $f^{\rm Match}(q_T)$ that is unity in the $q_T/\mStar\rightarrow0$ limit
and vanishes the asymptotic-resummed difference at $q_T\gtrsim\mStar$:
\begin{equation}
 f^{\rm Match}(q_T) = \frac{1}{1+(q_T/q_{T}^{\rm Match})^4}.
 \label{eq:cssMatchFn}
\end{equation}
We set $q_{T}^{\rm Match} = \mStar/3$~\cite{Han:1991sa}, and write the matched, triply differential distribution~\cite{Arnold:1990yk,Han:1991sa}
\begin{eqnarray}
  \frac{d\sigma^{\rm Matched}}{{d\mStar^2 ~dy ~dq_T^2}} &=&  \frac{d\sigma^{\rm FO}}{{d\mStar^2 ~dy ~dq_T^2}} +  
f^{\rm Match}(q_T)
 \left[
 \frac{d\sigma^{\rm Resum}}{{d\mStar^2 ~dy ~dq_T^2}} - \frac{d\sigma^{\rm Asymp}}{{d\mStar^2 ~dy ~dq_T^2}}
 \right].
 \label{eq:comboFORes}
\end{eqnarray}
In the $q_T/\mStar\rightarrow0$ limit, the longitudinal momentum fractions of initial-state partons $q,\overline{q}'$ are 
\begin{equation}
 \xi_1 = e^y \frac{\mStar}{\sqrt{s}}, \quad \xi_2 = e^{-y} \frac{\mStar}{\sqrt{s}}, \quad d\mStar^2 ~dy = s ~d\xi_1 ~d\xi_2.
\end{equation}
Though untrue for the resummed and asymptotic pieces at large $q_T/\mStar$, their combined contribution vanishes by construction.
For the FO calculation in this limit, 
$\mStar\rightarrow\sqrt{\hat{s}}$ and the above momentum fractions are for initial-state partons $q,\overline{q}',$ and $g$.
With this, one may write
\begin{eqnarray}
\frac{d\sigma^{\rm \Delta}}{d\mStar^2 ~dy ~dq_T^2} \equiv 
\frac{d\sigma^{\rm Resum}}{d\mStar^2 ~dy ~dq_T^2} - \frac{d\sigma^{\rm Asymp}}{d\mStar^2 ~dy ~dq_T^2} 
= \frac{1}{s}\left(\frac{d\sigma^{\rm Resum}}{d\xi_1 ~d\xi_2 ~dq_T^2} - \frac{d\sigma^{\rm Asymp}}{d\xi_1 ~d\xi_2 ~dq_T^2}\right).
\end{eqnarray}
Defining $d\sigma^\Delta/d\xi_1 d\xi_2 dq_T$ as the quantity in the parentheses, one at last has
\begin{eqnarray}
  \frac{d\sigma^{\rm Matched}}{dq_T} &=&  
  \frac{d\sigma^{\rm FO}}{dq_T} 
  +
  \int_{\tau_0}^1 d\xi_1 \int^1_{\tau_0/\xi_1}d\xi_2 ~
  \frac{d\sigma^{\rm \Delta}}{d\xi_1 ~d\xi_2 ~dq_T}, \quad \tau_0 = \frac{4m_T^2}{s}.\label{eq:cssMatchInt}
\end{eqnarray}
To account for the appropriate normalization at $\mathcal{O}(\alpha_s)$, 
the area under the above distribution is scaled to equal $\sigma^{NLO}$ in Eq.~(\ref{eq:SigmaNLO}). 
The resummed and asymptotic terms are now discussed.

\subsection{Recoil $q_T$ Resummation}
Recalling the resummed expression, but in terms of $\xi_i$,
\begin{eqnarray}
 \frac{d\sigma^{\rm Resum}}{d\xi_1 ~d\xi_2 ~dq_T^2} &=& 
 \frac{1}{4}
 \int^\infty_0 db^2 ~J_0(b q_T) ~\times W ~\times ~\hat{\sigma}^{B}(q\overline{q} \rightarrow T ~\overline{T}),\label{eq:cssMaster}
 \\
  W &=& e^{-S_{NP}(b,\mStar,\xi_1,\xi_2)} ~ e^{-S_P(b_*,\mStar)} ~  \tilde{W}(b_*,\xi_1,\xi_2).
\end{eqnarray}
Technically, $W$ is ill-defined for $b>1/\Lambda_{\rm QCD}$ due to divergent $\alpha_s(b^{-2})$.
This is ameliorated~\cite{Collins:1984kg,Davies:1984sp} by $S_{NP}$, 
which is valid for all $b$ and approaches unity when $b\ll 1/\Lambda_{\rm QCD}$, and by introducing
\begin{equation}
 b_{*}(b) = \frac{b}{\sqrt{1 + b^2 / b_{\max}^2}} \leq b_{\max}, \quad 1/b_{\max} > \Lambda_{\rm QCD}.
\end{equation}
This approach, however, requires that $S_{NP}$ be extracted from data and dependent on $b_{\max}$.

We choose the non-perturbative Sudakov factor given by the BLNY parameterization~\cite{Landry:1999an,Landry:2002ix}
\begin{equation}
{S}_{NP}(b,Q,\xi_1,\xi_2) = b^2\left[ g_1 + g_2\log\left(\frac{Q}{2Q_{\rm BLNY}}\right) + g_1 g_3 \log(100 \xi_1 \xi_2) \right].
 \label{eq:cssSudakovNP}
\end{equation}
Fits for $g_i$ to Tevatron $Z$ boson data with $Q_{\rm BLNY} = 1.6\GeV$ and $b_{\max}=0.5\GeV^{-1}$ give~\cite{Landry:2002ix}
\begin{equation}
 g_1 = 0.21^{+0.01}_{-0.01}\GeV^{2}, \quad  g_2 = 0.68^{+0.01}_{-0.02}\GeV^{2}, \quad  g_3 = -0.60^{+0.05}_{-0.04}.
\end{equation}
To numerically integrate over the domain of $b$, we exploit the $b$-dependence of $S_{NP}$ and define: 
\begin{equation}
 h(b) \equiv e^{-b^2 g_1}, \quad \tilde{S}_{NP} \equiv S_{NP} - b^2 g_1 =
 b^2\left[ g_2\log\left(\frac{Q}{2Q_{\rm BLNY}}\right) + g_1 g_3 \log(100 \xi_A \xi_B) \right].
 \label{eq:cssChangeOfV1}
\end{equation}
The $b$ integral now takes the manageable form
\begin{equation}
 \int_0^\infty db^2 ~e^{-S_{NP}}  = \frac{1}{g_1}\int^1_0 dh ~e^{-\tilde{S}_{NP}}.
\end{equation}

The perturbative Sudakov-like form factor is given by
\begin{equation}
 S_P(b_*,Q) = \int^{Q^2}_{c_0^2/b^2_{*}}\frac{d\mu^2}{\mu^2}\left[A\log\frac{Q^2}{\mu^2} + B\right],
 \label{eq:cssSudakovPFull}
\end{equation}
where the low-scale integration limit is 
$\mu = c_0/b_*$, with $c_0 = 2e^{-\gamma_E}$ and $\gamma_E\approx 0.577$ is the Euler-Mascheroni constant.
The functions $A$ and $B$ can be expanded perturbatively in powers of $\alpha_s(\mu^2)$. 
For a resummed calculation at LL, we take $A$ and $B$ to $\mathcal{O}(\alpha_s^{n=1})$. 
The expressions are~\cite{Collins:1984kg}
\begin{eqnarray}
 A = \sum_{n=1}^\infty A^{(n)}\left(\frac{\alpha_s}{2\pi}\right)^n &=& A^{(1)}\left(\frac{\alpha_s}{2\pi}\right) + \mathcal{O}(\alpha_s^2),
 \quad A^{(1)} = 2 C_F, 
\label{eq:cssAexpand}
 \\
 B = \sum_{n=1}^\infty B^{(n)}\left(\frac{\alpha_s}{2\pi}\right)^n &=& B^{(1)}\left(\frac{\alpha_s}{2\pi}\right)+ \mathcal{O}(\alpha_s^2),
 \quad B^{(1)} = -3 C_F.
\label{eq:cssBexpand}
 \end{eqnarray}
At one-loop, $\alpha_s$ is given by
\begin{equation}
 \alpha_s(\mu^2) = \frac{\alpha_s(M_Z^2)}{1 + b_0 ~\alpha_s(M_Z^2)~ \log\left(\cfrac{\mu^2}{M_Z^2}\right)}, 
 \quad b_0 = \frac{1}{12\pi}\left(11 N_c - 2n_f\right),
\end{equation}
and allows Eq.~(\ref{eq:cssSudakovPFull}) to be evaluated analytically~\cite{Dreiner:2006sv}. The result is
\begin{eqnarray}
 S_P(b_*,Q) &\approx& 
 \int^{Q^2}_{c_0^2/b^2_{*}}\frac{d\mu^2}{\mu^2}~C_F\frac{\alpha_s(\mu^2)}{2\pi}\left[2\log\frac{Q^2}{\mu^2} -3\right]
 \\
 &=&
 \frac{C_F}{2\pi b_0^2\alpha_s(M_Z^2)}
 \left[2+ 2 b_0 \alpha_s(M_Z^2)t(Q^2) -3 b_0 \alpha_s(M_Z^2) \right]
 \log X
 \nonumber\\
 & & -\cfrac{C_F}{\pi b_0}\log\left(\cfrac{b_*^2 Q^2}{c_0^2}\right) ,  \label{eq:cssSudakovP}
 \\
 X &=& \frac{1+ b_0 \alpha_s(M_Z^2)t(Q^2)}{1+ b_0 \alpha_s(M_Z^2)t(c_0^2/b_*^2)},
 \quad  
 t(\mu^2) = \log\frac{\mu^2}{M_Z^2}.
\end{eqnarray}

In the perturbative limit $b\ll1/\Lambda_{\rm QCD}$, 
the TMD $\mathcal{F}^T$ factorize into universal Wilson coefficient functions for $i\rightarrow j$ splitting, $C_{ji},$ 
and the usual transverse momentum-independent PDFs:
\begin{equation}
 \mathcal{F}^{T}_{q/p}(\xi_1,b^2,\mu_f^2) \mathcal{F}^{T}_{\overline{q}'/p}(\xi_2,b^2,\mu_f^2) =
 \sum_{i,k=q,\overline{q}',g} \left[C_{qi}\otimes f_{i/p}\right](\xi_1,\mu_f^2) \times \left[C_{\overline{q}'k}\otimes f_{k/p}\right](\xi_2,\mu_f^2).
 \label{eq:tmpDecomp}
\end{equation}
The summation is over all partons $i,k$ that can split into $q,\overline{q}'$. The convolution notation denotes
\begin{equation}
 \left[C_{ji}\otimes f_{i/p}\right](\xi,\mu_f^2) \equiv \int^1_\xi \frac{dz}{z} ~ C_{ji}(z)~f_{i/p}\left(\frac{\xi}{z},\mu_f^2\right)
\end{equation}
Like $A,B$ in $S_{P}$, $C_{ji}$ can be expanded in powers of $\alpha_s$ and we take $C_{ji}$ to $\mathcal{O}(\alpha_s^{n=0})$ ~\cite{Collins:1984kg}:
\begin{equation}
 C_{ji}(z) = \sum_{n=0}^\infty C^{(n)}_{ji}(z)\left(\frac{\alpha_s}{2\pi}\right)^n \approx C^{(0)}_{ji}(z) = \delta_{ji}\delta(1-z).
\end{equation}
Formally, $C_{ji}(z)^{(1)}$ is an $\mathcal{O}(\alpha_s^2)$ contribution except at $q_T=0$, where it $\mathcal{O}(\alpha_s)$.
Hence, our $q_T$ spectrum is accurate to $\mathcal{O}(\alpha_s)$ in shape everywhere but the origin~\cite{Han:1991sa},
an acceptable error considering the uncertainty in our knowledge of $m_T$.
Evaluating the convolutions, the product of TMDs is
\begin{equation}
  \mathcal{F}^{T}_{q/p}(\xi_1,b^2,\mu_f^2) \mathcal{F}^{T}_{\overline{q}/p}(\xi_2,b^2,\mu_f^2)
 \approx \left[C_{qq}^{(0)}\otimes f_{q/p}\right]\times\left[C_{\overline{q}\overline{q}}^{(0)}\otimes f_{\overline{q}/p}\right]
 = f_{q/p}(\xi_1,\mu_f^2)f_{\overline{q}/p}(\xi_2,\mu_f^2).
 \label{eq:cssTMDApprox}
\end{equation}
Equating the collinear and impact scales, $\mu^2_f=c_0^2/b_*^{2}$, $\tilde{W}(b_*,\xi_1,\xi_2)$ becomes
\begin{equation}
  \tilde{W}(b_*,\xi_1,\xi_2) = \sum_{q,\overline{q}}
  \left[
  f_{q/p}\left(\xi_1,\frac{c_0^2}{b_*^{2}}\right)f_{\overline{q}/p}\left(\xi_2,\frac{c_0^2}{b_*^{2}}\right)
  +
  f_{q/p}\left(\xi_2,\frac{c_0^2}{b_*^{2}}\right)f_{\overline{q}/p}\left(\xi_1,\frac{c_0^2}{b_*^{2}}\right)
 \right].  
 \label{eq:cssResumW}
\end{equation}
Though valid only for $q_T/\mStar\rightarrow0$, 
extension to $b\gtrsim \Lambda_{\rm QCD}$ is allowed by $S_{NP}$ and $J_0(x)$, which vanish for large $b$ and $x=q_T b$.
After simplification, the resummed expression to LL accuracy is
\begin{equation}
   \cfrac{d\sigma^{\rm Resum}_{LL}}{d\xi_1 ~d\xi_2 ~dq_T} = 
  \cfrac{q_T}{2g_1} \int^1_0 dh ~J_0(b q_T) 
  ~e^{-\tilde{S}_{NP}(b,\mStar,\xi_1,\xi_2)} 
  ~e^{-{S}_P(b_*,\mStar)} ~
     \tilde{W}(b_*,\xi_1,\xi_2) \times  \hat{\sigma}^{B},
     \label{eq:cssLL}
\end{equation}
where expressions for the relevant factors are given in Eqs.~(\ref{eq:cssChangeOfV1}),~(\ref{eq:cssSudakovP}), and (\ref{eq:cssResumW}).
In essence, the expression preceding $\hat{\sigma}^{B}$ is the unintegrated TMD parton luminosity for DY systems.

\subsection{Asymptotic Expansion of Resummed Expression}
The asymptotic piece is obtained by formally expanding the resummed expression in powers of $\alpha_s$, keeping only terms as singular as $1/q_T^2$,
and evaluating the impact parameter integral. The result is 
\begin{eqnarray}
  &\cfrac{d\sigma^{\rm Asymp}}{d\xi_1 ~d\xi_2 ~dq_T}& = 
  \frac{1}{q_T}
  \frac{\alpha_s(\mu^2)}{\pi}
  \sum_{q,\overline{q}} ~
  \Bigg[
  \left(A\log\frac{\mStar^2}{q_T^2} + B\right)f_{q/p}(\xi_1,\mu^{2})f_{\overline{q}/p}(\xi_2,\mu^{2})
    \nonumber\\
  &+& 
  \bar{f}_{q/p}(\xi_1,\mu^{2})f_{\overline{q}/p}(\xi_2,\mu^{2})+ 
  f_{q/p}(\xi_1,\mu^{2})\bar{f}_{\overline{q}/p}(\xi_2,\mu^{2})+ 
(\xi_1\leftrightarrow\xi_2)\Bigg] \times \hat{\sigma}^B.\label{eq:cssAsymp}
\end{eqnarray}
The renormalization and factorization scales here must be set equal to the FO scales, i.e., Eq.~(\ref{eq:renfactScale}),
in order to avoid spurious logarithms containing ratios of factorization/renormalization scales.
To $\mathcal{O}(\alpha_s),$ $A$ and $B$ are given in Eqs.~(\ref{eq:cssAexpand})-(\ref{eq:cssBexpand}).
The PDFs, corrected for single parton splitting,  are 
\begin{eqnarray}
 \bar{f}_{q/p}(\xi,\mu^2) &=& 
 \frac{\alpha_s(\mu^2)}{2\pi}\sum_{i=q,g}  \left[P_{qi}\otimes f_{i/p}\right]\left(\xi,\mu^2\right)
  \\
  &=& 
 \frac{\alpha_s(\mu^2)}{2\pi}\left[
 \int^1_\xi \frac{dz}{z} ~\left[P_{qq}(z)\right]_+ f_{q/p}\left(\frac{\xi}{z},\mu^2\right)
 +
 \int^1_\xi \frac{dz}{z} ~P_{qg}(z) f_{g/p}\left(\frac{\xi}{z},\mu^2\right) 
  \right],
\end{eqnarray}
where the AP splitting functions are in Eq.~(\ref{eq:pssAP}) and 
the ``plus distribution'' $[P(z)]_+$ is defined  as 
\begin{eqnarray}
 \int_0^1  dz ~[P(z)]_{+} ~f(z,\mu^2) &\equiv& \int_0^1   dz ~P(z) ~[f(z,\mu^2) - f(1)], \quad\text{and}\\
 \int_\xi^1 dz~[P(z)]_{+} ~f(z,\mu^2) &\equiv& \int_\xi^1 dz ~P(z) ~[f(z,\mu^2) - f(1)] + f(1)\int_0^\xi dz ~P(z).
\end{eqnarray}
Expanding the $q\rightarrow q$ integral under the plus distribution gives the expression
\begin{eqnarray}
  \int^1_\xi \frac{dz}{z} ~\left[P_{qq}(z)\right]_+ f_{q/p}\left(\frac{\xi}{z},\mu^2\right)
&=&
\int^1_\xi dz ~
\left[ \frac{1}{z}  P_{qq}(z)~f_{q/p}\left(\frac{\xi}{z},\mu^2\right) - 
 \frac{2C_F}{(1-z)}f_{q/p}(\xi,\mu^2)\right]
 \nonumber\\
 & & ~+~ 2\left[1+C_F\log(1-\xi)\right]~f_{q/p}(\xi,\mu^2),
\end{eqnarray}
and allows one to write
 \begin{eqnarray}
 \bar{f}_{q/p}(\xi,\mu_f^2) &=& 
 \frac{\alpha_s(\mu^2)}{2\pi}
\int^1_\xi \frac{dz}z ~
\left[  
	P_{qq}(z)~f_{q/p}\left(\frac{\xi}{z},\mu^2\right) 
	-\frac{2C_Fz}{(1-z)}f_{q/p}(\xi,\mu^2)
	+P_{qg}(z) f_{g/p}\left(\frac{\xi}{z},\mu^2\right) 
\right]
   \nonumber\\
 &  & +  \frac{\alpha_s(\mu^2)}{\pi} \left[1+C_F\log(1-\xi)\right]~f_{q/p}(\xi,\mu^2),
\end{eqnarray}
There is slight technical distinction between $\bar{f}$ here and $\tilde{f}$ in Eq.~(\ref{eq:pssSplitPDF}), 
where hard collinear splittings in PSS are addressed.
The $\bar{f}$ are regulated by subtracting out individual singular points via the plus distribution,
whereas $\tilde{f}$ are regulated by the soft cutoff $\delta_S$.

\end{appendix}

\hrulefill


\begin{thebibliography}{99}
\bibitem{Ma:1998dn} 
  E.~Ma,
  \textit{Pathways to naturally small neutrino masses,}
  Phys.\ Rev.\ Lett.\  {\bf 81}, 1171 (1998)
  [hep-ph/9805219].

\bibitem{Minkowski:1977sc} 
  P.~Minkowski,
  \textit{$\mu \to e\gamma$ at a Rate of One Out of $10^{9}$ Muon Decays?,}
  Phys.\ Lett.\ B {\bf 67}, 421 (1977).
  
\bibitem{Mohapatra:1979ia} 
  R.~N.~Mohapatra and G.~Senjanovic,
  \textit{Neutrino Mass and Spontaneous Parity Violation,}
  Phys.\ Rev.\ Lett.\  {\bf 44}, 912 (1980).
  
  \bibitem{Yanagida:1979as} 
  T.~Yanagida,
  \textit{Horizontal Symmetry And Masses Of Neutrinos,}
  Conf.\ Proc.\ C {\bf 7902131}, 95 (1979).
  
  \bibitem{GellMann:1980vs} 
  M.~Gell-Mann, P.~Ramond and R.~Slansky,
  \textit{Complex Spinors and Unified Theories,}
  Conf.\ Proc.\ C {\bf 790927}, 315 (1979)
  [arXiv:1306.4669 [hep-th]].
  
  \bibitem{Schechter:1980gr} 
  J.~Schechter and J.~W.~F.~Valle,
  \textit{Neutrino Masses in SU(2) x U(1) Theories,}
  Phys.\ Rev.\ D {\bf 22}, 2227 (1980).
  
  \bibitem{Shrock:1980ct} 
  R.~E.~Shrock,
  \textit{General Theory of Weak Leptonic and Semileptonic Decays. 1. Leptonic Pseudoscalar Meson Decays, with Associated Tests For, and Bounds on, Neutrino Masses and Lepton Mixing,}
  Phys.\ Rev.\ D {\bf 24}, 1232 (1981).
        
\bibitem{Magg:1980ut} 
  M.~Magg and C.~Wetterich,
  \textit{Neutrino Mass Problem and Gauge Hierarchy,}
  Phys.\ Lett.\ B {\bf 94}, 61 (1980).
  
  \bibitem{Cheng:1980qt} 
  T.~P.~Cheng and L.~F.~Li,
  \textit{Neutrino Masses, Mixings and Oscillations in SU(2) x U(1) Models of Electroweak Interactions,}
  Phys.\ Rev.\ D {\bf 22}, 2860 (1980).
  
  \bibitem{Lazarides:1980nt} 
  G.~Lazarides, Q.~Shafi and C.~Wetterich,
  \textit{Proton Lifetime and Fermion Masses in an SO(10) Model,}
  Nucl.\ Phys.\ B {\bf 181}, 287 (1981).
  
  \bibitem{Mohapatra:1980yp} 
  R.~N.~Mohapatra and G.~Senjanovic,
  \textit{Neutrino Masses and Mixings in Gauge Models with Spontaneous Parity Violation,}
  Phys.\ Rev.\ D {\bf 23}, 165 (1981).
   
  \bibitem{Foot:1988aq} 
  R.~Foot, H.~Lew, X.~G.~He and G.~C.~Joshi,
  \textit{Seesaw Neutrino Masses Induced by a Triplet of Leptons,}
  Z.\ Phys.\ C {\bf 44}, 441 (1989).
  
   
\bibitem{Barger:2003qi} 
  V.~Barger, D.~Marfatia and K.~Whisnant,
  \textit{Progress in the physics of massive neutrinos,}
  Int.\ J.\ Mod.\ Phys.\ E {\bf 12}, 569 (2003)
  [hep-ph/0308123].
  
  \bibitem{Atre:2009rg} 
  A.~Atre, T.~Han, S.~Pascoli and B.~Zhang,
  \textit{The Search for Heavy Majorana Neutrinos,}
  JHEP {\bf 0905}, 030 (2009)
  [arXiv:0901.3589 [hep-ph]].
  
\bibitem{Chen:2011de} 
  M.~C.~Chen and J.~Huang,
  \textit{TeV Scale Models of Neutrino Masses and Their Phenomenology,}
  Mod.\ Phys.\ Lett.\ A {\bf 26}, 1147 (2011)
  [arXiv:1105.3188 [hep-ph]].
  
\bibitem{Deppisch:2015qwa} 
  F.~F.~Deppisch, P.~S.~B.~Dev and A.~Pilaftsis,
  \textit{Neutrinos and Collider Physics,}
  arXiv:1502.06541 [hep-ph].
  
  
 
\bibitem{Keung:1983uu} 
  W.~Y.~Keung and G.~Senjanovic,
  \textit{Majorana Neutrinos and the Production of the Right-handed Charged Gauge Boson,}
  Phys.\ Rev.\ Lett.\  {\bf 50}, 1427 (1983).
  
  \bibitem{Pilaftsis:1991ug} 
  A.~Pilaftsis,
  \textit{Radiatively induced neutrino masses and large Higgs neutrino couplings in the standard model with Majorana fields,}
  Z.\ Phys.\ C {\bf 55}, 275 (1992)
  [hep-ph/9901206].
  
\bibitem{Datta:1993nm} 
  A.~Datta, M.~Guchait and A.~Pilaftsis,
  \textit{Probing lepton number violation via majorana neutrinos at hadron supercolliders,}
  Phys.\ Rev.\ D {\bf 50}, 3195 (1994)
  [hep-ph/9311257].
  
\bibitem{Han:2006ip} 
  T.~Han and B.~Zhang,
  \textit{Signatures for Majorana neutrinos at hadron colliders,}
  Phys.\ Rev.\ Lett.\  {\bf 97}, 171804 (2006)
  [hep-ph/0604064].
  



\bibitem{Rizzo:1983zz} 
  T.~G.~Rizzo,
  \textit{Doubly Charged Higgs Bosons and Lepton Number Violating Processes,}
  Phys.\ Rev.\ D {\bf 25}, 1355 (1982)
  [Phys.\ Rev.\ D {\bf 27}, 657 (1983)].
   
\bibitem{Gunion:1996pq} 
  J.~F.~Gunion, C.~Loomis and K.~T.~Pitts,
  \textit{Searching for doubly charged Higgs bosons at future colliders,}
  eConf C {\bf 960625}, LTH096 (1996)
  [hep-ph/9610237].
  

  
  \bibitem{Rizzo:1981dm} 
  T.~G.~Rizzo and G.~Senjanovic,
  \textit{Grand Unification and Parity Restoration at Low-Energies. 1. Phenomenology,}
  Phys.\ Rev.\ D {\bf 24}, 704 (1981)
  [Phys.\ Rev.\ D {\bf 25}, 1447 (1982)].
  
  
\bibitem{Dev:2013wba} 
  P.~S.~B.~Dev, A.~Pilaftsis and U.~-k.~Yang,
  \textit{New Production Mechanism for Heavy Neutrinos at the LHC,}
  arXiv:1308.2209 [hep-ph].

  \bibitem{Alva:2014gxa} 
  D.~Alva, T.~Han and R.~Ruiz,
  \textit{Heavy Majorana neutrinos from $W\gamma$ fusion at hadron colliders,}
  JHEP {\bf 1502}, 072 (2015)
  [arXiv:1411.7305 [hep-ph]].
  

  \bibitem{Ruiz:2015gsa} 
  R.~E.~Ruiz, Pp. 109-110,
  \textit{Hadron Collider Tests of Neutrino Mass-Generating Mechanisms},  Ph.D. thesis, University of Pittsburgh (2015)
  [arXiv:1509.06375 [hep-ph]].
  
\bibitem{Muhlleitner:2003me} 
  M.~Muhlleitner and M.~Spira,
  \textit{A Note on doubly charged Higgs pair production at hadron colliders,}
  Phys.\ Rev.\ D {\bf 68}, 117701 (2003)
  [hep-ph/0305288].
  
    \bibitem{Sullivan:2002jt} 
  Z.~Sullivan,
  \textit{Fully differential $W^\prime$ production and decay at next-to-leading order in QCD,}
  Phys.\ Rev.\ D {\bf 66}, 075011 (2002)
  [hep-ph/0207290].
  
  \bibitem{Jezo:2014wra} 
  T.~Jezo, M.~Klasen, D.~R.~Lamprea, F.~Lyonnet and I.~Schienbein,
  \textit{NLO+NLL limits on $W'$ and $Z'$ gauge boson masses in general extensions of the Standard Model,}
  JHEP {\bf 1412}, 092 (2014)
  [arXiv:1410.4692 [hep-ph]].
  
  \bibitem{Gavin:2010az} 
  R.~Gavin, Y.~Li, F.~Petriello and S.~Quackenbush,
  \textit{FEWZ 2.0: A code for hadronic Z production at next-to-next-to-leading order,}
  Comput.\ Phys.\ Commun.\  {\bf 182}, 2388 (2011)
  [arXiv:1011.3540 [hep-ph]].
  
  \bibitem{Gavin:2012sy} 
  R.~Gavin, Y.~Li, F.~Petriello and S.~Quackenbush,
  \textit{W Physics at the LHC with FEWZ 2.1,}
  Comput.\ Phys.\ Commun.\  {\bf 184}, 208 (2013)
  [arXiv:1201.5896 [hep-ph]].

\bibitem{Binosi:2003yf} 
  D.~Binosi and L.~Theussl,
  \textit{JaxoDraw: A Graphical user interface for drawing Feynman diagrams,}
  Comput.\ Phys.\ Commun.\  {\bf 161}, 76 (2004)
  [hep-ph/0309015].
  
    \bibitem{Fabricius:1981sx} 
  K.~Fabricius, I.~Schmitt, G.~Kramer and G.~Schierholz,
  \textit{Higher Order Perturbative QCD Calculation of Jet Cross-Sections in e+ e- Annihilation,}
  Z.\ Phys.\ C {\bf 11}, 315 (1981).
  
  \bibitem{Kramer:1986mc} 
  G.~Kramer and B.~Lampe,
  \textit{Jet Cross-Sections in e+ e- Annihilation,}
  Fortsch.\ Phys.\  {\bf 37}, 161 (1989).
  
  \bibitem{Baer:1989jg} 
  H.~Baer, J.~Ohnemus and J.~F.~Owens,
  \textit{A Next-To-Leading Logarithm Calculation of Jet Photoproduction,}
  Phys.\ Rev.\ D {\bf 40}, 2844 (1989).
  
  \bibitem{Harris:2001sx} 
  B.~W.~Harris and J.~F.~Owens,
  \textit{The Two cutoff phase space slicing method,}
  Phys.\ Rev.\ D {\bf 65}, 094032 (2002)
  [hep-ph/0102128].
  
  
\bibitem{Collins:1981uk}
  J.~C.~Collins and D.~E.~Soper,
  \textit{Back-To-Back Jets in QCD,}
  Nucl.\ Phys.\ B {\bf 193} (1981) 381
   [Nucl.\ Phys.\ B {\bf 213} (1983) 545].

  \bibitem{Collins:1981va} 
  J.~C.~Collins and D.~E.~Soper,
  \textit{Back-To-Back Jets: Fourier Transform from B to K-Transverse,}
  Nucl.\ Phys.\ B {\bf 197}, 446 (1982).

\bibitem{Collins:1984kg} 
  J.~C.~Collins, D.~E.~Soper and G.~F.~Sterman,
  \textit{Transverse Momentum Distribution in Drell-Yan Pair and W and Z Boson Production,}
  Nucl.\ Phys.\ B {\bf 250}, 199 (1985).
  
  
  
 \bibitem{Franceschini:2008pz} 
  R.~Franceschini, T.~Hambye and A.~Strumia,
  \textit{Type-III see-saw at LHC,}
  Phys.\ Rev.\ D {\bf 78}, 033002 (2008)
  [arXiv:0805.1613 [hep-ph]].

  \bibitem{delAguila:2008cj} 
  F.~del Aguila and J.~A.~Aguilar-Saavedra,
  \textit{Distinguishing seesaw models at LHC with multi-lepton signals,}
  Nucl.\ Phys.\ B {\bf 813}, 22 (2009)
  [arXiv:0808.2468 [hep-ph]].

    \bibitem{Arhrib:2009mz} 
  A.~Arhrib, B.~Bajc, D.~K.~Ghosh, T.~Han, G.~Y.~Huang, I.~Puljak and G.~Senjanovic,
  \textit{Collider Signatures for Heavy Lepton Triplet in Type I+III Seesaw,}
  Phys.\ Rev.\ D {\bf 82}, 053004 (2010)
  [arXiv:0904.2390 [hep-ph]].
  
  \bibitem{AguilarSaavedra:2009ik} 
  J.~A.~Aguilar-Saavedra,
  \textit{Heavy lepton pair production at LHC: Model discrimination with multi-lepton signals,}
  Nucl.\ Phys.\ B {\bf 828}, 289 (2010)
  [arXiv:0905.2221 [hep-ph]].
  
  \bibitem{Li:2009mw} 
  T.~Li and X.~G.~He,
  \textit{Neutrino Masses and Heavy Triplet Leptons at the LHC: Testability of Type III Seesaw,}
  Phys.\ Rev.\ D {\bf 80}, 093003 (2009)
  [arXiv:0907.4193 [hep-ph]].
  
  
\bibitem{Bandyopadhyay:2010wp} 
  P.~Bandyopadhyay and E.~J.~Chun,
  \textit{Displaced Higgs production in type III Seesaw,}
  JHEP {\bf 1011}, 006 (2010)
  [arXiv:1007.2281 [hep-ph]].
  
\bibitem{CMS:2015mza} 
  CMS Collaboration [CMS Collaboration],
  \textit{Search for Heavy Lepton Partners of Neutrinos in pp Collisions at 8 TeV, in the Context of Type III Seesaw Mechanism,}
  CMS-PAS-EXO-14-001.
 
 \bibitem{Aad:2015cxa} 
  G.~Aad {\it et al.} [ATLAS Collaboration],
  \textit{Search for type-III Seesaw heavy leptons in $pp$ collisions at $\sqrt{s}= 8$ TeV with the ATLAS Detector,}
  Phys.\ Rev.\ D {\bf 92}, no. 3, 032001 (2015)
  [arXiv:1506.01839 [hep-ex]].
   
 \bibitem{Pierce:1993gj} 
  D.~Pierce and A.~Papadopoulos,
  \textit{Radiative corrections to neutralino and chargino masses in the minimal supersymmetric model,}
  Phys.\ Rev.\ D {\bf 50}, 565 (1994)
  [hep-ph/9312248].
   
\bibitem{Ibe:2006de} 
  M.~Ibe, T.~Moroi and T.~T.~Yanagida,
  \textit{Possible Signals of Wino LSP at the Large Hadron Collider,}
  Phys.\ Lett.\ B {\bf 644}, 355 (2007)
  [hep-ph/0610277].
     
 \bibitem{Baer:1997nh} 
  H.~Baer, B.~W.~Harris and M.~H.~Reno,
  \textit{Next-to-leading order slepton pair production at hadron colliders,}
  Phys.\ Rev.\ D {\bf 57}, 5871 (1998)
  [hep-ph/9712315].
  
 \bibitem{Arnold:1990yk} 
  P.~B.~Arnold and R.~P.~Kauffman,
  \textit{W and Z production at next-to-leading order: From large q(t) to small,}
  Nucl.\ Phys.\ B {\bf 349}, 381 (1991).
  
    \bibitem{Han:1991sa} 
  T.~Han, R.~Meng and J.~Ohnemus,
  \textit{Transverse momentum distribution of Z boson pairs at hadron supercolliders,}
  Nucl.\ Phys.\ B {\bf 384}, 59 (1992).
  
  
  
\bibitem{Hahn:2004fe} 
  T.~Hahn,
  \textit{CUBA: A Library for multidimensional numerical integration},
  Comput.\ Phys.\ Commun.\  {\bf 168}, 78 (2005)
  [hep-ph/0404043].
  
    \bibitem{Alwall:2006yp} 
  J.~Alwall {\it et al.},
  \textit{A Standard format for Les Houches event files,}
  Comput.\ Phys.\ Commun.\  {\bf 176}, 300 (2007)
  [hep-ph/0609017].
  
    \bibitem{Alloul:2013bka} 
  A.~Alloul, N.~D.~Christensen, C.~Degrande, C.~Duhr and B.~Fuks,
  \textit{FeynRules  2.0 - A complete toolbox for tree-level phenomenology,}
  Comput.\ Phys.\ Commun.\  {\bf 185}, 2250 (2014)
  [arXiv:1310.1921 [hep-ph]].
  
  \bibitem{Christensen:2008py} 
  N.~D.~Christensen and C.~Duhr,
  \textit{FeynRules - Feynman rules made easy,}
  Comput.\ Phys.\ Commun.\  {\bf 180}, 1614 (2009)
  [arXiv:0806.4194 [hep-ph]].
  
 \bibitem{Alwall:2014hca} 
  J.~Alwall, R.~Frederix, S.~Frixione, V.~Hirschi, F.~Maltoni, O.~Mattelaer, H.-S.~Shao and T.~Stelzer {\it et al.},
  \textit{The automated computation of tree-level and next-to-leading order differential cross sections, and their matching to parton shower simulations,}
  JHEP {\bf 1407}, 079 (2014)
  [arXiv:1405.0301 [hep-ph]].
  
  \bibitem{Beringer:1900zz} 
  J.~Beringer {\it et al.}  [Particle Data Group Collaboration],
  \textit{Review of Particle Physics (RPP),}
  Phys.\ Rev.\ D {\bf 86}, 010001 (2012).
  
\bibitem{Owens:2012bv} 
  J.~F.~Owens, A.~Accardi and W.~Melnitchouk,
  \textit{Global parton distributions with nuclear and finite-$Q^2$ corrections,}
  Phys.\ Rev.\ D {\bf 87}, no. 9, 094012 (2013)
  [arXiv:1212.1702 [hep-ph]].
  
  \bibitem{Brock:2014tja} 
  R.~Brock {\it et al.},
  \textit{Planning the Future of U.S. Particle Physics (Snowmass 2013): Chapter 3: Energy Frontier,}
  arXiv:1401.6081 [hep-ex].
  
  
  
  
\bibitem{Altarelli:1979ub} 
  G.~Altarelli, R.~K.~Ellis and G.~Martinelli,
  \textit{Large Perturbative Corrections to the Drell-Yan Process in QCD,}
  Nucl.\ Phys.\ B {\bf 157}, 461 (1979).
  
    \bibitem{Hinchliffe:1988ap} 
  I.~Hinchliffe and S.~F.~Novaes,
  \textit{On the Mean Transverse Momentum of Higgs Bosons at the {SSC},}
  Phys.\ Rev.\ D {\bf 38}, 3475 (1988).

  \bibitem{Kauffman:1991jt} 
  R.~P.~Kauffman,
  \textit{Higgs boson p(T) in gluon fusion,}
  Phys.\ Rev.\ D {\bf 44}, 1415 (1991).
   
\bibitem{Davies:1984sp} 
  C.~T.~H.~Davies, B.~R.~Webber and W.~J.~Stirling,
  \textit{Drell-Yan Cross-Sections at Small Transverse Momentum,}
  Nucl.\ Phys.\ B {\bf 256}, 413 (1985).
  
      \bibitem{Landry:1999an} 
  F.~Landry, R.~Brock, G.~Ladinsky and C.~P.~Yuan,
  \textit{New fits for the nonperturbative parameters in the CSS resummation formalism,}
  Phys.\ Rev.\ D {\bf 63}, 013004 (2001)
  [hep-ph/9905391].
  
    \bibitem{Landry:2002ix} 
  F.~Landry, R.~Brock, P.~M.~Nadolsky and C.~P.~Yuan,
  \textit{Tevatron Run-1 $Z$ boson data and Collins-Soper-Sterman resummation formalism,}
  Phys.\ Rev.\ D {\bf 67}, 073016 (2003)
  [hep-ph/0212159].
  
  \bibitem{Dreiner:2006sv} 
  H.~K.~Dreiner, S.~Grab, M.~Kramer and M.~K.~Trenkel,
  \textit{Supersymmetric NLO QCD corrections to resonant slepton production and signals at the Tevatron and the CERN LHC,}
  Phys.\ Rev.\ D {\bf 75}, 035003 (2007)
  [hep-ph/0611195].
\end{thebibliography}
\end{document}